\documentclass[conference,compsoc]{IEEEtran}
\ifCLASSOPTIONcompsoc
  \usepackage[nocompress]{cite}
\else
  \usepackage{cite}
\fi
\ifCLASSINFOpdf
\else
\fi
\usepackage{graphicx}
\usepackage{hyperref}
\usepackage{url}
\usepackage{wrapfig}
\usepackage{subcaption}
\usepackage{caption}
\usepackage{algorithm}
\usepackage{algpseudocode}
\usepackage{amsmath}
\usepackage{multirow}
\usepackage{amsfonts}
\usepackage[table,xcdraw]{xcolor}
\usepackage{tcolorbox}
\usepackage{balance}

\usepackage[normalem]{ulem}

\begin{document}
%
\title{GenBFA: An Evolutionary Optimization Approach to Bit-Flip Attacks on LLMs
}

\author{\large Sanjay Das$^{1*}$, Swastik Bhattacharya$^1$, Souvik Kundu$^2$, Shamik Kundu$^2$, Anand Menon$^{1}$, \\Arnab Raha$^2$ and Kanad Basu$^1$\\
\textit{$^1$University of Texas at Dallas, USA}; \textit{$^2$Intel Corporation, USA}\\
$^*$Corresponding author: Sanjay Das (Email: \textit{sanjay.das@utdallas.edu})}
\maketitle

\begin{abstract}


Large language models (LLMs) have transformed natural language processing (NLP) by excelling in tasks like text generation and summarization. Their adoption in mission-critical applications highlights an emerging concern: LLMs' vulnerability to hardware-based threats, particularly bit-flip attacks (BFAs). BFAs involve fault injection methods, including techniques such as Rowhammer, which target the model parameters stored in memory, thereby undermining model integrity and performance. Given the vast parameter space of LLMs, identifying critical parameters for BFAs is challenging and often infeasible. Existing research suggests that transformer-based architectures are inherently more robust to BFAs compared to traditional deep neural networks (DNNs). However, we challenge this assumption for the first time, demonstrating that as few as three bit-flips can cause a performance collapse in an LLM with billions of parameters.

Current BFA techniques fail to exploit this vulnerability due to the difficulty of traversing the large parameter space to identify critical parameters. To address this, we propose AttentionBreaker, a novel framework specifically designed for LLMs that facilitates efficient parameter-space traversal to identify critical parameters. Additionally, we introduce GenBFA, an evolutionary optimization strategy, to further narrow down the critical parameter set and identify the most vulnerable bits, facilitating a more efficient attack. Empirical findings indicate the pronounced vulnerability of LLMs to AttentionBreaker: for instance, perturbing merely $3$ bits (\(4.129 \times 10^{-9}\%\) of total parameters) in LLaMA3-8B 8-bit weight quantized (W8) model results in total performance collapse, with accuracy on Massive Multitask Language Understanding (MMLU) tasks plummeting from \(67.3\%\) to \(0\%\) and perplexity escalating from \(12.6\) to \(4.72\times10^5\). These results highlight the framework's efficacy in exploiting and exposing inherent vulnerabilities in LLM architectures. 
Code is open
sourced at: \url{https://github.com/TIES-Lab/attnbreaker}.
\end{abstract}

\IEEEpeerreviewmaketitle

\section{Introduction}
\label{sec:intro}
The rise in popularity of LLMs has fundamentally expanded the capabilities of Artificial Intelligence (AI), demonstrating remarkable proficiency in generating human-like text, interpreting nuanced context, and executing complex reasoning tasks. These advancements have not only reshaped natural language processing but have also extended AI applications into diverse fields such as computer vision and scientific research, heralding a new era of AI-driven solutions~\cite{radford2018improving, chang2024survey, xu2024survey}. Given their pervasive use, it is critical to analyze the vulnerability profile of LLMs against both software-based and hardware-based threats to ensure their secure and reliable deployment ~\cite{das2024security}. 
\begin{figure}[t]
    \centering
    \includegraphics[width=0.9\linewidth]{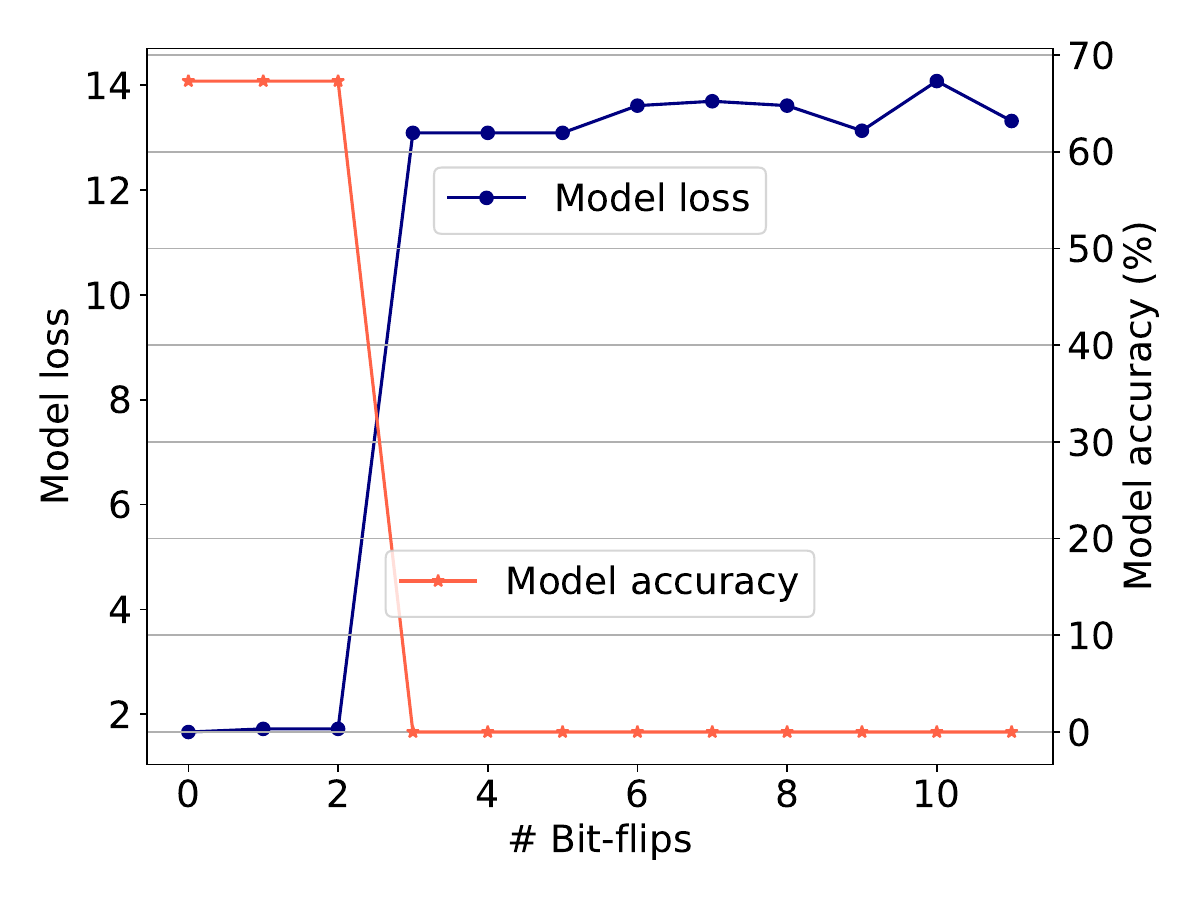}
    \caption{Bit-flip attack with AttentionBreaker on LLaMA3-8B-Instruct W8 model on MMLU benchmark.}
    \label{fig:AttBreaker}
\end{figure}
Of particular concern are hardware threats like BFAs, which exploit hardware vulnerabilities to corrupt memory regions that store the model's weight parameters, thereby compromising model integrity and performance~\cite{rakin2019bit, qian2023survey, kundubit}. BFAs employ fault injection attacks such as DeepHammer~\cite{yao2020deephammer} to manipulate specific bits within DRAM, altering essential model weights to degrade functionality. Despite advancements in-memory technology, recent techniques enable remote, non-physical memory manipulation, perpetuating the threat landscape for BFAs~\cite{hayashi2011non, shuvo2023comprehensive}. While BFAs have been extensively studied in the context of DNNs~\cite{qian2023survey, rakin2019bit, chen2021proflip}, their implications for transformer-based architectures, including LLMs, remain largely unexplored, highlighting a crucial gap in current research. Addressing this gap is critical for evaluating transformer models' susceptibility to bit-flip vulnerabilities and ensuring their reliability, especially in mission-critical applications.


Existing research on BFAs targeting Transformer-based models suggests that these architectures exhibit greater inherent resilience compared to traditional DNNs. This robustness is largely attributed to the unique structural properties of Transformer models~\cite{nazari2024forget}. Specifically, studies indicate that the inherent redundancy and modularity of Transformers facilitate internal error compensation across different model components, allowing these architectures to mitigate localized faults more effectively than DNNs~\cite{bai2021transformers}. 
In this work, we challenge the presumed robustness of Transformer models by demonstrating, for the first time, that as few as three bit-flips in an 8-billion parameter model—comprising tens of billions of bits—can result in catastrophic degradation, reducing the model's accuracy to 0\% on standard benchmarks such as MMLU tasks (refer Figure \ref{fig:AttBreaker}). This finding emphasizes the extreme vulnerability of LLMs to BFAs. 
However, existing BFA methods are unable to exploit this vulnerability, primarily due to the vast parameter space of LLMs, which complicates the identification of critical parameters for attack. 
\emph{To address this limitation, we propose a novel, systematic approach to BFAs specifically tailored to uncover and exploit bit-flip vulnerabilities in LLMs.
This represents the first application of such a method, achieving unprecedented effectiveness in targeting these models.} Figure \ref{fig:bfatransformer} illustrates an example of a BFA on a transformer-based model, demonstrating how a single bit-flip in the Query weight matrix propagates through the attention process, leading to distributed errors that ultimately distort the model’s output. 

To execute an adversarial attack on LLMs, the main challenge is navigating the extremely large parameter space to identify critical parameters whose perturbation significantly degrades model performance. Given the nonlinear nature of this task, heuristic optimization techniques present a suitable approach. With this premise, we propose a novel algorithm based on evolutionary strategies~\cite{lambora2019genetic}, designed to identify a minimal yet critical subset of parameters for an efficient BFA. \emph{{This approach leverages principles of natural selection and genetic variation to iteratively refine the parameter set, enhancing optimization efficiency.}} Our contributions are summarized below:
\begin{figure}[t]
    \centering
    \includegraphics[width=1.0\linewidth]{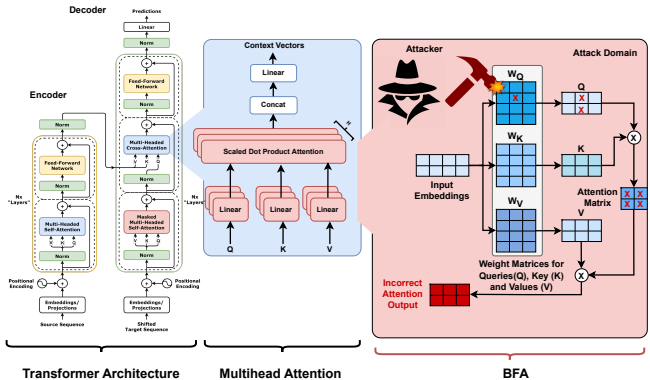}
    \caption{Bit-flip attack on transformer-based architecture.}
    \label{fig:bfatransformer}
\end{figure}



\begin{itemize}

\item For the first time, we challenge the notion of LLMs' high resilience to BFAs, demonstrating that as few as three bit-flips can critically degrade model performance, even for LLMs with billions of parameters and tens of billions of bits.

\item To navigate the vast parameter space of LLMs, often cited as a reason for their assumed resilience, we introduce a novel framework, \textbf{AttentionBreaker}, enabling efficient and targeted bit-flip attacks.

\item We present a novel evolutionary algorithm, \textbf{GenBFA}, to identify the most critical bits in LLMs, optimizing the selection of targets for bit-flip attacks.






\item Our framework uncovers the significant vulnerability of LLMs, as evidenced by a mere \textbf{three} bit-flips—a flip rate of just $\mathbf{4.129 \times 10^{-9}}\%$—in LLaMA3-8B, reduces MMLU accuracy from 67.3\% to 0\% and raises perplexity from 12.62 to \(4.72 \times 10^5\). 


\end{itemize}

The rest of the paper is organized as follows: Section \ref{sec:background} provides relevant background information. Section \ref{sec:variables} introduces the variable notations used throughout the work. In Section \ref{sec:motivation}, we present the rationale for this investigation. The proposed methodology is detailed in Section \ref{sec:methodology}. Section \ref{sec:results} outlines the experimental setup and discusses the results. An ablation study examining framework parameter values is presented in Section \ref{sec:ablation}. Concluding remarks are offered in Section \ref{sec:conclusion}.

\section{Background}\label{sec:background}

This section provides essential background on the Transformer architecture and Bit-Flip Attacks (BFAs), offering the necessary context to understand the interplay between Transformer models and potential vulnerabilities.
\subsection{Transformer Models}

The Transformer architecture is foundational for LLMs and Vision Transformers, driving advancements in NLP and image classification~\cite{vaswaniattention, radford2018improving, chang2024survey, xu2024survey}. 
It comprises two main components: the encoder to convert input data into a structured intermediate representation for improved feature understanding and the decoder to generate output sequences from this encoded information\cite{zhang2023dive}. There are three task-specific configurations: Encoder-only models for input-driven tasks (e.g., text classification, named entity recognition), Decoder-only models for generative tasks, and Encoder-decoder models for tasks requiring input-dependent generation (e.g., machine translation, summarization).

A critical innovation in the Transformer is the attention mechanism, particularly multi-head attention, which prioritizes relevant input elements~\cite{zhang2023dive}. Attention weights are computed to emphasize informative tokens, while masks can exclude specific tokens, such as padding.

In each multi-head attention layer, the model derives three parameters for each input token: \textbf{Query} (\(Q\)), \textbf{Key} (\(K\)), and \textbf{Value} (\(V\)), calculated through linear transformations of input vectors. Attention scores assess token importance across the sequence. Each head processes distinct \(Q\), \(K\), and \(V\) subsets, facilitating parallel analysis of diverse token relationships. The individual attention outputs are concatenated and linearly combined to produce the final attention output, enhancing the model's ability to capture complex dependencies and improve task performance in both NLP and vision domains.
\subsection{Bit-flip Attack (BFA)}
The robustness, security, and safety of modern computing systems are intrinsically linked to memory isolation, which is enforced through both software and hardware mechanisms~\cite{frassetto2018imix}. However, due to read-disturbance phenomena, contemporary DRAM chips remain susceptible to memory isolation breaches~\cite{6853210}. RowHammer~\cite{mutlu2019rowhammer} is a well-documented example where repeated DRAM row (hammering) access induces bitflips in adjacent rows due to electrical interference. 


Techniques for inducing bit-flips have enabled BFAs, designed to degrade model accuracy through strategic memory faults. BFAs are designed to minimize the number of bit-flips required, enhancing stealth and reducing detection risk. 
DeepHammer~\cite{yao2020deephammer} exemplifies advanced BFAs by employing Progressive Bit Search (PBS)~\cite{rakin2019bit}, a gradient-based algorithm to locate vulnerable bits. In each iteration, DeepHammer selects \(n\) vulnerable bits by ranking model parameters by gradient sensitivity. Each bit in this subset is individually flipped, forming a corresponding loss set \(L\). This process iterates across layers, yielding \(n \times l\) candidate bits and loss sets, with the most vulnerable bit identified as the one producing the highest loss. The bit-flipping continues until the targeted misclassification is achieved. Although limited to flipping a single bit per page, DeepHammer assumes an adversary can flip multiple bits per DRAM page.

\section{Variable Notations}\label{sec:variables}
In this section, we define the variable notations used throughout this paper, as summarized in Table \ref{table:notations}. 

\begin{table}[t]
\caption{Summary of Variables and Notations.}
\label{table:notations}
\centering
\renewcommand{\arraystretch}{1.1}
\resizebox{1\linewidth}{!}{\begin{tabular}{|c|c|}
\hline
\rowcolor[HTML]{C0C0C0} \textbf{Variable notation} & \textbf{Description} \\ \hline \hline
$\mathcal{M}$ & The model \\ \hline
$L = \{l_1,l_2, \cdots, l_n\}$ & Layers in a model \\ \hline
$\mathbf{W} = \{w_1, w_2, \dots, w_n\}$ & All model parameters \\ \hline
$\mathcal{L}(\mathbf{W})$ & The loss function based on the model parameters \\ \hline
$\nabla\mathcal{L}(\mathbf{W}) = \nabla \mathbf{W} $ & Gradient of the loss with respect to parameters \\ \hline
$\mathbf{S} = \{s_1, s_2,\dots, s_n\}$ & Sensitivity of parameters to perturbations \\ \hline
$\alpha$ & Ratio of contribution to $\mathbf{S}$ by $\mathbf{W}$ and $\nabla\mathbf{W}$ \\ \hline
\texttt{cardinality} & Function to provide size of a given set \\ \hline
$\mathcal{R} = \{r_1,r_2, \cdots, r_n\}$ & Set of sampling rates \\ \hline
$I = \{i_1,i_2, \cdots, i_n\}$ & Set of indices to target for bit-flip perturbations \\ \hline
$\mathbf{W}_N, \nabla\mathbf{W}_N$ & Normalized model weights and gradients \\ \hline
\end{tabular}}
\end{table}

Consider in a model \(\mathcal{M}\) with \(L\)-layers, parameterized by \(\mathbf{W} = \{w_1, w_2, \dots, w_n\}\), \(\mathcal{L}(\mathbf{W})\) denotes the associated loss function. For simplicity, we denote the gradient of the loss function \(\nabla \mathcal{L}(\mathbf{W})\) as \(\nabla \mathbf{W}\) throughout this paper. To assess the sensitivity of model parameters, we define a sensitivity metric \(\mathbf{S}\), with \(\mathbf{S}_{\text{mag}}\) and \(\mathbf{S}_{\text{grad}}\) capturing sensitivity based on parameter magnitudes and gradients, respectively. To balance the influence of gradients and magnitudes in the computation of a hybrid sensitivity metric, \(\mathbf{S}\), a sensitivity ratio \(\alpha \in [0,1]\) is introduced. The attention weight matrix is represented by \(\mathbf{A}\), with corresponding query and key vectors \(q_i\) and \(k_j\), scaled by \(d_k\). 
The function \texttt{cardinality} is defined to evaluate the number of parameters in a given parameter set. Let \(w'\) represent a perturbed version of weight \(w\), and \(\mathcal{R}\) denote the set of sampling rates for parameter sampling during sensitivity analysis. The normalized gradients and weights are denoted by \(\nabla \mathbf{W}_N\) and \(\mathbf{W}_N\), respectively. Finally, \(I\) is the set of indices where bit flips are applied to perturb specific parameters.

\section{Motivation}\label{sec:motivation}


Fault attacks on DNNs typically exploit vulnerabilities in hardware implementations or manipulate parameters to induce misclassifications, which has been extensively explored in existing literature~\cite{rakin2019bit, kundubit}. However, the specific domain of fault injection attacks on LLMs, which involves introducing controlled parameter perturbations by memory corruptions to manipulate their output, remains underexplored~\cite{hector2022closer}. This uncharted territory presents a unique challenge due to the complex nature of transformer-based architectures. Fault injection attacks on LLMs could potentially result in unintended or biased text generation, misclassification, undermined coherence, or subtle manipulations that are difficult to detect.
To address this research gap, we systematically evaluate BFA impacts on transformer-based models, particularly LLMs, and propose a novel and highly efficient AttentionBreaker framework for LLMs.
\subsection{Proxy for Sensitivity}\label{sec:motivation_layersensitivity}
For performing BFA, it is essential to analyze model parameter sensitivity and vulnerability profiles independently of assumptions of robustness. 
In particular, parameters with larger gradients or higher magnitudes may exhibit amplified sensitivity, whereby perturbations yield disproportionately large effects on the output.  
In full-precision models (\textit{e.g.}, float32), where weights \( w \) span a large representable range \([-3.4 \times 10^{38}, 3.4 \times 10^{38}]\), even a single bit-flip can induce substantial perturbations, particularly for large-magnitude weights. For such weights, the sensitivity \( \mathbf{S}_{\text{mag}} \) can be approximated as:
\begin{equation}
    \mathbf{S}_{\text{mag}} \propto |\mathbf{W}|
\end{equation}
\begin{figure*}[t!]
  \centering
  \hfill
  \begin{subfigure}{0.32\linewidth}
    \centering
    \includegraphics[width=1.05\linewidth]{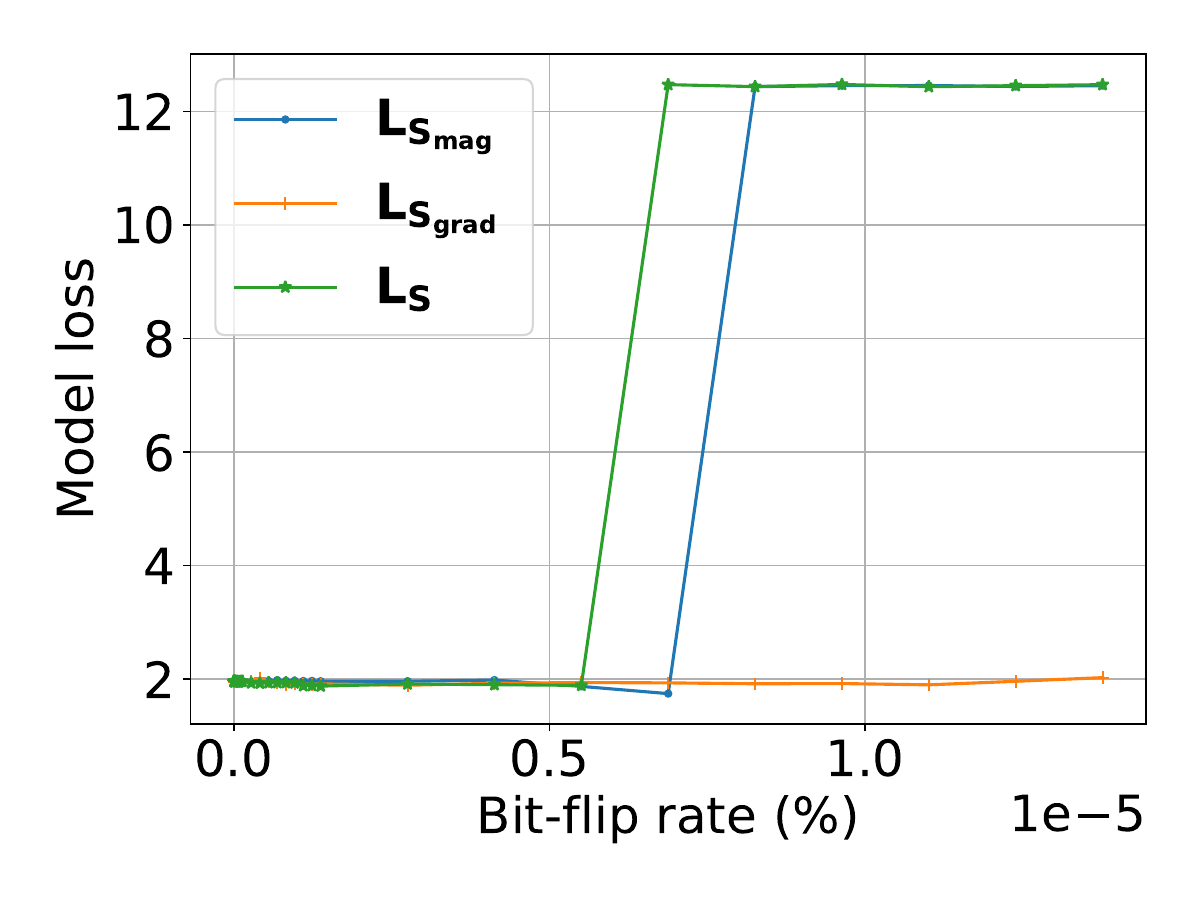}
    \caption{}
    \label{fig:layer_sens_mot}
  \end{subfigure}
  \begin{subfigure}{0.32\linewidth}
    \includegraphics[width=1.05\linewidth]{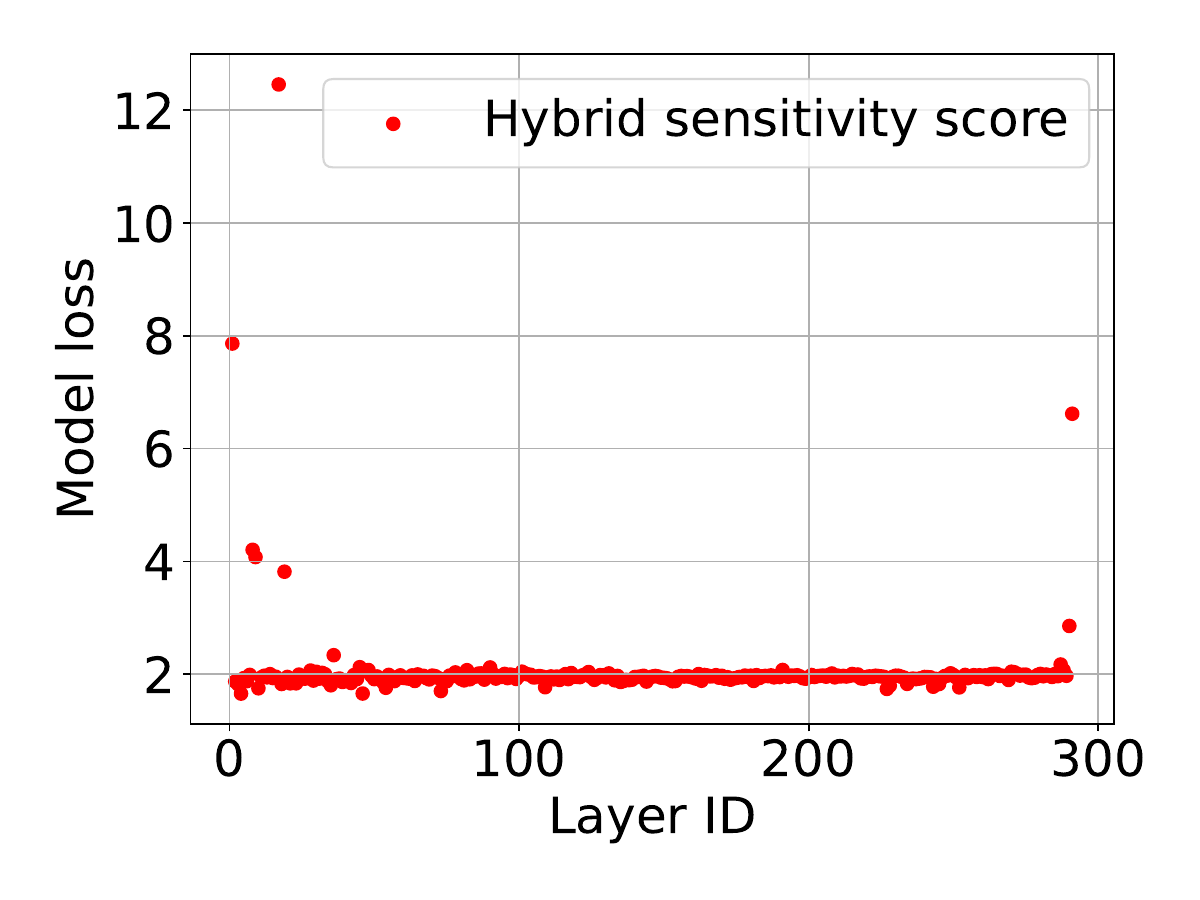}
    \caption{}
    \label{fig:parametersens}
  \end{subfigure}
  \hfill
  \begin{subfigure}{0.32\linewidth}
    \centering
    \includegraphics[width=1.05\linewidth]{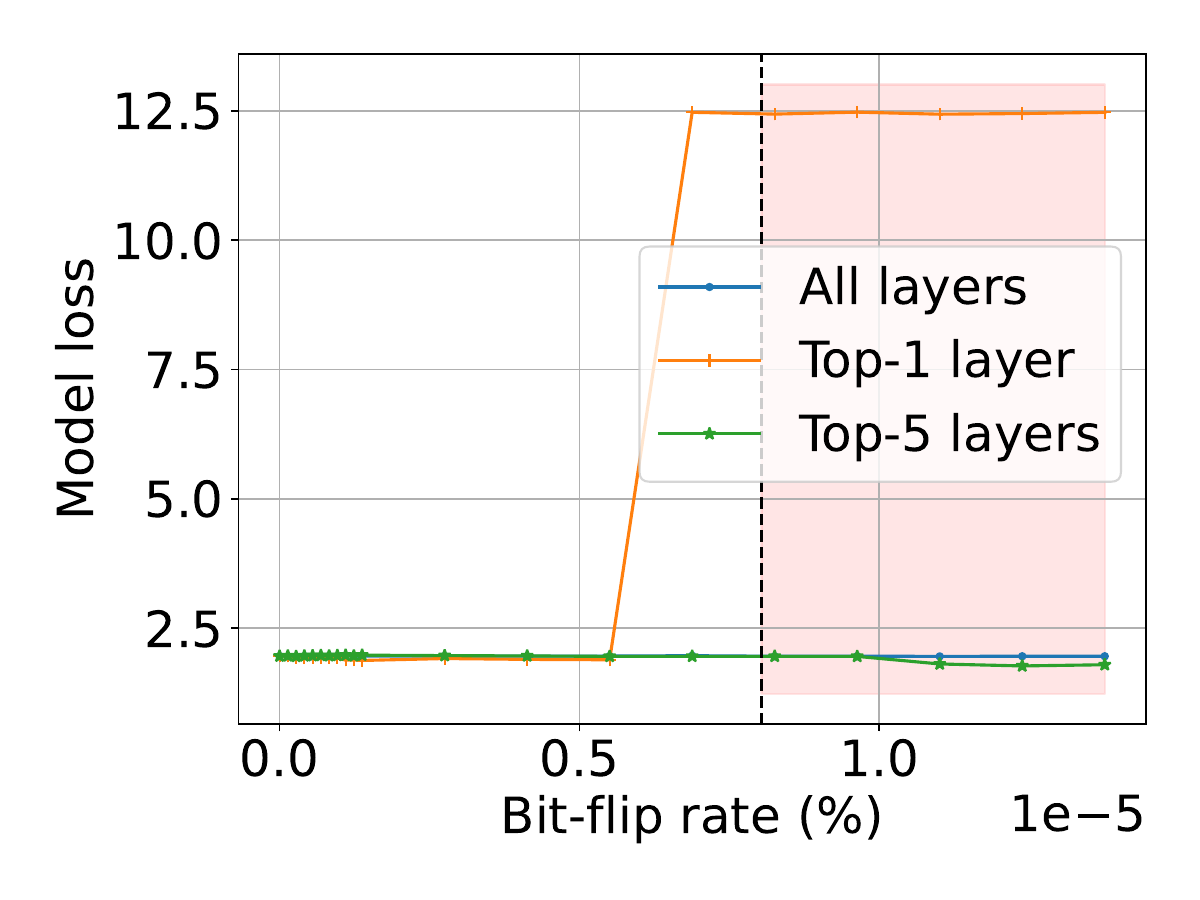}
    \caption{}
    \label{fig:subset_selection}
  \end{subfigure}
  \caption{LLaMA3-8B W8 (a) sensitivity proxy comparison and (b) layer sensitivity analysis and (c) weight subset reduction on the `Astronomy' task of MMLU benchmark.}
  \label{fig:main_motivation}
\end{figure*}

However, gradient-derived sensitivity \( \mathbf{S}_{\text{grad}} \) provides a complementary perspective, capturing the degree to which a weight change affects the model loss:
\begin{equation}
    \mathbf{S}_{\text{grad}} \propto |\nabla \mathbf{W}|
\end{equation}

This indicates that weights with large gradients are highly impactful on model performance, even if their magnitudes are small.
As precision decreases (\textit{e.g.}, from 16-bit floating point numbers to 8-bit and 4-bit integers, as well as binary and ternary formats), the perturbation impact from bit-flips diminishes. This is caused by the reduced range of representable values, which makes gradient-based sensitivity more significant. As a result, a hybrid sensitivity metric is required, one that considers both magnitude and gradient influences, and is therefore expressed as:
\begin{equation}
    \mathbf{S} \propto \alpha \cdot |\nabla \mathbf{W}| + (1 - \alpha) \cdot |\mathbf{W}|
\end{equation}

This formulation enables precise identification of high-impact parameters under varying precision, guiding robust sensitivity analysis and targeted parameter selection. By investigating the sensitivity metric, critical parameters in the model can be identified and perturbed with targeted bit-flips to achieve an adversarial attack. 





To analyze how this hybrid sensitivity score ($\mathbf{S}$) performs in quantifying the sensitivity of model parameters, we experiment with the LLaMA3-8B W8 model~\cite{dubey2024llama}\cite{dettmers2022llmint8}. This model was selected for its widely used precision level and moderate size, which render it representative of standard open-source LLMs. We select top-$k$ parameters in a single layer of the model based on gradients ($|\nabla \mathbf{W}|$), magnitudes ($|\mathbf{W}|$), and the hybrid sensitivity metric $\mathbf{S}$. Here, $\mathbf{S}$ is calculated by normalizing weights and gradients for accommodating variation in scales, as shown in Equation \ref{eq:sensitivity}. 

\begin{equation}
    \mathbf{S} = \alpha \cdot |\nabla \mathbf{W}_N| + (1 - \alpha) \cdot |\mathbf{W}_N|
    \label{eq:sensitivity}
\end{equation}
As per this equation, \(\alpha = 0\) emphasizes weight magnitudes, \(\alpha = 1\) emphasizes gradients, and intermediate values balance the two factors.
The selected parameters are then subjected to bit-flips, and the change in the model loss is plotted in Figure \ref{fig:layer_sens_mot}. We observe that using the hybrid metric $\mathbf{S}$, loss ($L_{\mathbf{S}}$) increases with less number of bit-flips than the other two metrics. This demonstrates the superiority of the hybrid metric in identifying critical model parameters. 

\subsection{Layer Sensitivity Analysis}\label{sec:mot_sensitivity}
Identifying critical bits in LLMs presents a formidable challenge due to the sheer volume of their parameter space. However, identifying a sensitive layer is considerably more manageable, as the number of layers is significantly lower than the total number of parameters. To quantify layer sensitivity, we choose top-\(r\%\) bit-flips in each layer, guided by the hybrid sensitivity score \(\mathbf{S}\), and observe the resulting model loss \(\mathcal{L}\) on the LLaMA3-8B W8 model. The losses vary significantly across layers, as illustrated in Figure \ref{fig:parametersens}. Several layers have very high sensitivity, while many are much more resilient. This observation indicates that by focusing only on layers with high sensitivity, we can substantially narrow the search space for identifying critical parameters.

\subsection{Weight Subset Reduction}\label{sec:weight_set_reduction}
In critical parameter search in LLMs, sensitive layers can be identified using previous analysis. However, this information alone is insufficient due to the vast parameter count within a single layer, which can range from 100,000 to 100 million. To address this, we perturb the top \(r\%\) of parameters in the top-$n$ layers in the model, where \(r\) or the sampling rate determines the number of bit-flips to be performed using a hybrid sensitivity metric \(\mathbf{S}\). The perturbation rate $r$ is varied from a very low value (0.0001) to considerably high (10), and the process continues until the perturbation-induced loss exceeds a predefined threshold. 

As illustrated in Figure \ref{fig:subset_selection}, perturbing only 0.8\% of weights in the most sensitive layer raises the loss to a predefined threshold of 8, leading to termination. Notably, this outcome is observed only when focusing on the most sensitive layer, not with other configurations. This result provides a subset of weights containing parameters critical to the model. By using this approach, the search space for identifying critical parameters can be significantly reduced.
\subsection{Weight Subset Optimization}\label{fig:subset_optimization}
The weight subset identified during the subset selection analysis in Section \ref{sec:weight_set_reduction} can be sufficiently large to render an effective adversarial attack impractical. For example, the subset size in Figure \ref{fig:subset_selection} is 5872, which is prohibitively large for efficient bit-flip attacks. This necessitates further optimization to narrow the subset to a much smaller set of weights and/or bits that enable a feasible attack. However, solving this problem through exhaustive approaches is computationally expensive due to the vast solution space.  
To address this challenge, we propose a heuristic optimization technique, leveraging the efficiency of explorative search strategies to quickly identify good-enough solutions. Specifically, we implement an optimization method inspired by evolutionary strategies, which simultaneously optimizes for higher model loss (to maximize attack effectiveness) and fewer bit-flips (to enhance attack efficiency). Using this approach, we successfully reduced the required bit-flips \textbf{from 5872 to 3}, achieving approximately $\mathbf{2\times10^3}$ reduction.  
This result highlights the effectiveness of the proposed method in drastically narrowing the search space, identifying a sub-hundred set of highly critical bits, and exposing their vulnerability to adversarial attacks.

\section{Proposed Attack Methodology}\label{sec:methodology}
Here, we describe the threat model, articulating the standard assumptions regarding the adversary’s capabilities for carrying out BFAs. Subsequently, we introduce an innovative attack framework, AttentionBreaker, designed to identify and manipulate the most critical bits within LLMs, precipitating a substantial degradation in model performance.
\begin{figure}[t]
    \centering
    \includegraphics[width=0.9\linewidth]{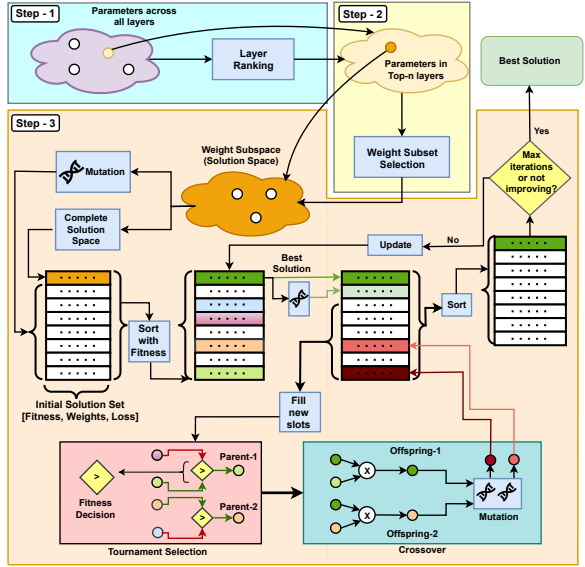}
    \caption{AttentionBreaker framework overview.}
    \label{fig:AttentionBreaker}
\end{figure}
\subsection{Threat Model}
In Machine Learning as a Service (MLaaS), deploying LLMs on shared computational platforms introduces significant security vulnerabilities. These platforms, while providing enhanced computational resources, expose LLMs to potential BFAs due to shared hardware components such as last-level caches and main memory. Attackers, even without explicit permissions to access user data, can exploit side channels to induce bit-flips in the memory cells, storing the LLM's parameters~\cite{shuvo2023comprehensive}. This threat model categorizes risks into three levels, defined by the attacker’s knowledge and degree of access to the model parameters: (1) At the basic level, attackers possess limited knowledge, allowing only minor perturbations that reduce model performance. (2) At the intermediate level, attackers gain targeted access to specific bits, enabling significant accuracy degradation through targeted manipulation. (3) At the advanced level, attackers have comprehensive access and sophisticated techniques, permitting them to induce severe outcomes such as complete model corruption or the embedding of backdoors.

In the context of our work, an attacker with elevated privileges gains unauthorized access to the memory where the LLM's weights are stored~\cite{liu2017fault}. The attacker aims to flip the least number of bits in the model to achieve the most adversarial impact. By exploiting fault injection techniques, such as the RowHammer attack~\cite{mutlu2019rowhammer}, the attacker selectively flips critical bits in the model's weights. 
This alteration can cause minute to severe model performance degradation, thus undermining the model's reliability, especially in mission-critical applications, including healthcare, finance, and autonomous systems.

\subsection{Proposed Attack Framework}
\begin{figure}[t]
    \centering
    \includegraphics[width=0.9\linewidth]{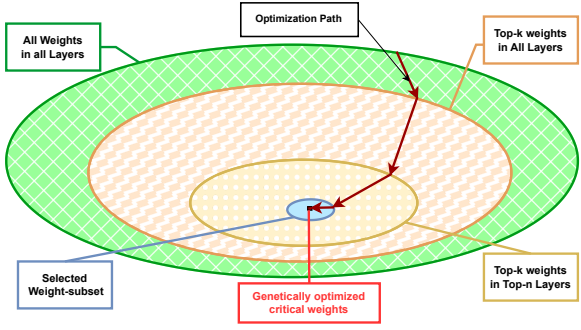}
    \caption{Step-wise critical weights optimization.}
    \label{fig:venn}
\end{figure}

Here, we explain in detail the proposed AttentionBreaker framework, which comprises three structured steps: (1) layer ranking, which generates a sensitivity profile for each layer; (2) weight subset selection, which extracts a critical subset of weights from this profile; and (3) weight subset optimization, which refines this subset to identify the model’s most vulnerable weights for an efficient attack. Figure \ref{fig:AttentionBreaker} depicts the overview of the proposed framework for the identification of critical parameters in the model. Figure \ref{fig:venn} illustrates how the critical weights are identified through a search space reduction and optimization process.
\subsubsection{Layer Ranking}\label{sec:layer_sensitivity}
For ranking LLM layers, we assess layer sensitivity as discussed in Section \ref{sec:mot_sensitivity}. We begin by quantifying the sensitivity of each layer’s parameters. For this purpose, each weight \( w_i \) in a layer is evaluated through a sensitivity scoring function that incorporates both its magnitude and gradient information using min-max normalization, as defined in Equation \ref{eq:sensitivity}.
Using these sensitivity scores, weights are aggregated and ranked in descending order to form a prioritized list of the most critical weights in each layer. The top-\( k \) weights within each layer are then identified for bit-flip perturbations, with \( k \) determined by Equation \ref{eq:topk}:

\begin{equation}\label{eq:topk}
    k = \texttt{cardinality}(\mathbf{W}^{(l)}) \times \frac{r}{100}
\end{equation}

To maximize the impact of bit-flips, perturbations are applied by targeting the most significant bits (MSBs) to induce the highest deviation for each selected weight. 
After applying these perturbations, we compute the model’s post-perturbation loss, \(\mathcal{L}^{(l)}\), which reflects the sensitivity of the layer. Note that a higher model loss indicates lower model performance, resulting in reduced accuracy in both text generation and classification tasks.
Iterating over each layer and recording the resulting perturbation losses generates a layer-wise bit-flip sensitivity profile of the model. This profile enables the identification of the most sensitive layers, guiding efficient analysis by prioritizing layers where perturbations most significantly degrade model performance. The layers are then ranked based on their corresponding perturbation loss.


\begin{algorithm}[t]
\caption{Layer Ranking}
\begin{algorithmic}[1]
\Require An  $L$-layer model $\mathcal{M}$ parameterized by $\mathbf{W}$, gradient $\nabla \mathbf{W}$, $\alpha$, sub-sampling $\%$  $r$ 
\State \text{Initialize sensitivity loss:} $\mathcal{L}_{sens} = [\text{ }]$
\Function{\texttt{SScore}}{$|\mathbf{W}|$, $|\nabla \mathbf{W}|$, $\alpha$}
    \State $|\mathbf{W}_N| \gets \texttt{Normalize}(|\mathbf{W}|)$
    \State $|\nabla \mathbf{W}_N| \gets \texttt{Normalize}(|\nabla \mathbf{W}|)$
    \State $\mathbf{S} \gets \alpha \cdot |\nabla \mathbf{W}_N| + (1 - \alpha) \cdot |\mathbf{W}_N|$
    \State \Return $\mathbf{S}$
\EndFunction
\Function{\texttt{BFLIP}}{$\mathbf{W}^{(l)}$, $pos$, $I$}
        \State $\mathbf{W}^{(l)}_{orig} \gets \mathbf{W}^{(l)}$
        \ForAll{$i \in I$}
        \State $w^{(l)}_i \gets w^{(l)}_i \oplus (1 \ll pos)$
    \EndFor
    \State $\mathcal{M} \gets \texttt{updateModel}(\mathcal{M}, \mathbf{W}^{(l)})$
    \State $\mathcal{L} \leftarrow \texttt{ComputeLoss}(\mathcal{M})$
    \State $\mathcal{M} \gets \texttt{updateModel}(\mathcal{M}, \mathbf{W}^{(l)}_{orig})$
    \State \Return $\mathcal{L}$
\EndFunction

\ForAll{$l \in L$}
    \State $k \gets r*\texttt{cardinality}(\mathbf{W}^{(l)})/100$
    \State $\textbf{S}^{(l)}$$\gets$ \texttt{SScore}($|\mathbf{W}^{(l)}|$, $|\nabla \mathbf{W}^{(l)}|$, $\alpha$)
    \State $I^{(l)}_{hybrid} \gets \texttt{Top$k$Index}(\mathbf{S}^{(l)})$
    
    \State $\mathcal{L}^{(l)} \leftarrow \texttt{BFLIP}({\mathbf{W}^{(l)}, pos, I^{(l)}_{hybrid}})$
    \State \text{Push} $\mathcal{L}^{(l)}$ \text{to} $\mathcal{L}_{sens}$

\EndFor    
    \State  $\mathcal{L}_{sens} \gets \texttt{SORT}(\mathcal{L}_{sens})$
\Ensure $\mathcal{L}_{sens}$
\end{algorithmic}
\label{alg:layer_ranking}
\end{algorithm}

Algorithm \ref{alg:layer_ranking} presents a systematic approach for conducting layer sensitivity analysis. The algorithm begins by initializing an empty set, $\mathcal{L}_{sens}$, designated to store the computed model loss resulting from parameter perturbations in each layer (\textit{line 1}).
The algorithm defines the function $\texttt{SSCORE}$, responsible for calculating sensitivity scores for each model parameter (\textit{lines 2-7}). Specifically, this function normalizes the weights and gradients to yield $\mathbf{W}_N$ and $\nabla \mathbf{W}_N$, respectively (\textit{lines 3-4}), and then computes the sensitivity scores $\mathbf{S}$ using Equation \ref{eq:sensitivity} (\textit{line 5}). Another function, $\texttt{BFLIP}$, is defined to calculate the model loss $\mathcal{L}$ when weight perturbations are applied to a specified layer (\textit{lines 8-17}). The function $\texttt{BFLIP}$ accepts as input the layer weight parameters $\mathbf{W}^{(l)}$, bit position $pos$ to be flipped, and perturbation indices $I$ (\textit{line 8}). It begins by storing a copy of the original weights $\mathbf{W}^{(l)}$ in $\mathbf{W}^{(l)}_{\text{orig}}$ for later restoration (\textit{line 9}). Then, it applies bit-flip perturbations to the weights $w_i^{(l)}$ at the specified indices $i \in I$ (\textit{lines 10-12}). The modified model, $\mathcal{M}$, is updated with the perturbed weights $\mathbf{W}^{(l)}$ via the $\texttt{UpdateModel}$ function (\textit{line 13}), after which the model loss corresponding to these perturbed weights is evaluated through $\texttt{ComputeLoss}$ (\textit{line 14}). To revert the perturbation, the function restores the model weights $\mathbf{W}^{(l)}$ to their original values $\mathbf{W}^{(l)}_{\text{orig}}$ (\textit{line 15}). Finally, the computed loss $\mathcal{L}$ is returned (\textit{line 16}).

The layer sensitivity analysis itself begins by iterating through each layer, assessing the model’s sensitivity to perturbations in each (\textit{lines 18-24}). First, the number of bits to be flipped, $k$, is calculated as a fraction of the total parameters in the layer, determined by $\texttt{cardinality}(\mathbf{W}^{(l)})$, multiplied by the sub-sampling rate $r$ (given as a percentage) and divided by 100 (\textit{line 19}). Sensitivity scores $\mathbf{S}^{(l)}$ for the weights in layer $l$ are then computed using the previously defined $\texttt{SSCORE}$ function (\textit{line 20}). Next, the $\texttt{Top-k}$ indices $I^{(l)}_{hybrid}$ are identified from $\mathbf{S}^{(l)}$ (\textit{line 21}). These indices, along with the weights $\mathbf{W}^{(l)}$ and bit position $pos$, are passed to the $\texttt{BFLIP}$ function to compute the perturbation-induced loss $\mathcal{L}^{(l)}_r$ (\textit{line 22}). The resulting loss $\mathcal{L}^{(l)}_r$ is then stored in $\mathcal{L}_{sens}$ (\textit{line 23}). This process continues until all layers have been processed, yielding the final sensitivity loss set $\mathcal{L}_{sens}$, which is then sorted and provided as output (\textit{line 25}).

\subsubsection{Weight Subset Selection}\label{sec:subset_selection}

\begin{algorithm}[t]
\caption{Weight Subset Selection}
\begin{algorithmic}[1]
\Require $\mathbf{W}$, $\nabla \mathbf{W}$, $\alpha$, $\mathcal{L}_{sens}$, \text{sub-sampling set }$\mathcal{R}$, \text{Loss threshold } $\mathcal{L}_{th}$, Number of top layers $n$
\State \text{Initialize sun-sampling loss:} $\mathcal{L}_{sub} = [\text{ }]$
\State $\mathcal{L}_{top} \gets \texttt{Top$n$}(\mathcal{L}_{sens})$
\ForAll{$l \in \mathcal{L}_{top}$}
\ForAll{$r \in \mathcal{R}$}
    \State $k \gets r*\texttt{cardinality}(\mathbf{W}^{(l)})/100$
    \State $\textbf{S}^{(l)}$$\gets$ \texttt{SScore}($|\mathbf{W}^{(l)}|$, $|\nabla \mathbf{W}^{(l)}|$, $\alpha$)
    \State $I^{(l)}_{hybrid} \gets \texttt{Top$k$Index}(\mathbf{S}^{(l)})$
    
    \State $\mathcal{L}^{(l)}_r \leftarrow \texttt{BFLIP}({\mathbf{W}^{(l)}, pos, I^{(l)}_{hybrid}})$
    \If{$\mathcal{L}^{(l)}_r \geq \mathcal{L}_{th}$} 
        \State \text{Push} $[l,\mathcal{L}^{(l)}_r,k,r]$ \text{to} $\mathcal{L}_{sub}$
        \State \textbf{break}
    \EndIf
    
\EndFor
\EndFor
\State $\mathcal{L}_{sub} \gets \texttt{SORT}(\mathcal{L}_{sub}) \text{ based on 3rd item }k$
\State $\mathcal{L}_{sub-1} \gets \mathcal{L}_{sub}[0]$
\State $l,k \gets \mathcal{L}_{sub-1}[0], \mathcal{L}_{sub-1}[2]$
\State $\textbf{S}^{(l)}$$\gets$ \texttt{SScore}($|\mathbf{W}^{(l)}|$, $|\nabla \mathbf{W}^{(l)}|$, $\alpha$)
\State $I^{(l)}_{hybrid} \gets \texttt{Top$k$Index}(\mathbf{S}^{(l)})$
\State $\mathbf{W}_{sub} \gets [l,I^{(l)}_{hybrid}]$
\Ensure $\mathbf{W}_{sub}$\Comment{Return weight subset}
\end{algorithmic}
\label{Algo:bitflip}
\end{algorithm}

To determine a reduced weight subset, we conduct a perturbation experiment as discussed in Section \ref{sec:weight_set_reduction}. We vary the sub-sampling rate \( r \) across a defined range to identify the minimum \( r \) value at which the perturbation-induced loss \( \mathcal{L} \) reaches or exceeds a predefined threshold, \( \mathcal{L}_{th} \). This threshold represents a critical degradation in model performance, indicating that further increments in \( r \) are unnecessary. 
By recording both \( r \) and the corresponding bit-flip count \( k \) for each layer, we can identify the smallest set of weights that meets \( \mathcal{L}_{th} \), enabling efficient subset selection.
Once each layer has been assessed for optimal sub-sampling values, we determine the specific layer \( l \) and corresponding values of \( k \) and \( r \) that yield the smallest subset satisfying the condition \( \mathcal{L} \geq \mathcal{L}_{th} \). The resulting subset $\mathbf{W}_{sub}$ represents the most efficient collection of weights for further refinement.

The weight subset selection procedure, outlined in Algorithm \ref{Algo:bitflip}, takes as inputs the model weights \(\mathbf{W}\), gradients \(\nabla \mathbf{W}\), sensitivity ratio \(\alpha\), layer sensitivity \(\mathcal{L}_{sens}\), sub-sampling rates \(\mathcal{R}\), a loss threshold \(\mathcal{L}_{th}\), and the number of top layers \(n\), and outputs a weight subset \(\mathbf{W}_{sub}\).

The process begins with initializing a list, \(\mathcal{L}_{sub}\), to record losses (\textit{line 1}). The top \(n\) layers are selected based on \(\mathcal{L}_{sens}\), derived in Section \ref{sec:layer_sensitivity} (\textit{line 2}). For each of these top layers, the algorithm iterates through sub-sampling ratios \(r\), computing the perturbation loss induced by bit flips (\textit{lines 3-14}). Specifically, the number of bits to flip, \(k\), is computed as a fraction of the total parameters in the layer using Equation \ref{eq:topk} (\textit{line 5}). Sensitivity scores \(\mathbf{S}^{(l)}\) are then calculated for layer \(l\) using the \texttt{SSCORE} function (\textit{line 6}), and the indices \(I^{(l)}_{hybrid}\) of the \texttt{Top-k} scores are identified (\textit{line 7}). These indices, along with the corresponding weights \(\mathbf{W}^{(l)}\) and a specified bit position \(pos\), are input to the \texttt{BFLIP} function, which returns the perturbation loss \(\mathcal{L}^{(l)}_r\) (\textit{line 8}).

If \(\mathcal{L}^{(l)}_r \geq \mathcal{L}_{th}\), this loss, along with layer index \(l\), \(k\), and \(r\), is stored in \(\mathcal{L}_{sub}\), and further iterations for that layer are halted (\textit{lines 9-12}); otherwise, iterations continue until \(\mathcal{R}\) is exhausted. After completing this sub-sampling procedure, the loss data in \(\mathcal{L}_{sub}\) is sorted by \(k\) (\textit{line 15}), yielding an ordered set where the smallest weight subset achieving the target loss is prioritized. The layer \(l\) and subset size \(k\) for this smallest subset are then extracted (\textit{lines 16-17}). Sensitivity scores \(\mathbf{S}^{(l)}\) are recalculated for layer \(l\) (\textit{line 18}), and the indices \(I^{(l)}_{hybrid}\) corresponding to the \texttt{Top-k} scores are selected (\textit{line 19}). Finally, the identified subset \(\mathbf{W}_{sub}\) for layer \(l\) and \(I^{(l)}_{hybrid}\) is returned as the algorithm's output.

\subsubsection{Weight-set Optimization - GenBFA}\label{sec:genetic}
After identifying the weight subset \(\mathbf{W}_{sub}\), further refinement is essential given that LLMs typically encompass billions of parameters and tens of billions of bits as discussed in Section \ref{fig:subset_optimization}. Consequently, an attack, even with the identified weight subset, may necessitate targeting thousands or even millions of parameters—implying tens of millions of bit-flips—which would be prohibitive from an attacker's perspective. Therefore, to address this issue, we systematically reduce \(\mathbf{W}_{sub}\) to $\mathbf{W}_{opt}$ ($\mathbf{W}_{opt} \subset \mathbf{W}_{sub}$) by isolating a smaller subset of parameters capable of achieving comparable degradation in model performance.
To address this objective, we design a novel algorithm inspired by evolutionary strategies, \textbf{GenBFA}, tailored for optimizing the parameter set to identify a minimal yet highly critical subset of weights. The GenBFA maximizes the impact-to-size ratio of the selected parameters, thereby enhancing the efficiency and effectiveness of the AttentionBreaker attack on LLMs.


\noindent\textbf{Objective Function and Fitness Evaluation:}
To formalize the weight subset optimization problem, let us define an initial weight subset $\mathbf{W}_{\text{sub}}$ and a target model loss threshold \(\mathcal{L}_{\text{th}}\). The objective is to identify a subset \(\mathbf{W}_{S_j} \subset \mathbf{W}_{\text{sub}}\) such that bit-flipping the weights in this subset maximizes the model's loss while minimizing the subset's \texttt{cardinality}, which reflects the number of bit-flips required to achieve a corresponding model loss, thus providing a measure of the attack's efficiency. Each candidate solution \(\mathbf{W}_{S_j} = \{ w_{j_1}, w_{j_2}, \ldots \}\) is therefore defined as a subset of \(\mathbf{W}_{\text{sub}}\).

For each candidate \(\mathbf{W}_{S_j}\), we compute a fitness value \(f(\mathbf{W}_{S_j})\) as follows. First, we define the sign function \(\operatorname{sgn}\left(\mathcal{L}_{S_j} - \mathcal{L}_{\text{th}}\right)\) based on whether the model loss \(\mathcal{L}_{S_j}\) (evaluated after bit-flipping the weights in \(\mathbf{W}_{S_j}\)) meets or exceeds the threshold \(\mathcal{L}_{\text{th}}\):

\begin{equation}
    \operatorname{sgn}\left(\mathcal{L}_{S_j} - \mathcal{L}_{\text{th}}\right) =
\begin{cases} 
+1, & \text{if } \mathcal{L}_{S_j} - \mathcal{L}_{\text{th}} \geq 0, \\ 
-1, & \text{if } \mathcal{L}_{S_j} - \mathcal{L}_{\text{th}} < 0.
\end{cases}
\end{equation}

Using this sign function, we define \(f(\mathbf{W}_{S_j})\) as:

\begin{equation}\label{eq:loss_fn}
    f(\mathbf{W}_{S_j}) = \operatorname{sgn}\left(\mathcal{L}_{S_j} - \mathcal{L}_{\text{th}}\right) \cdot \frac{\mathcal{L}_{S_j}}{\texttt{cardinality}(\mathbf{W}_{S_j})}
\end{equation}

Thus, this formulation maximizes model loss while minimizing the number of bit-flips, effectively balancing loss maximization with subset \texttt{cardinality}.

\noindent\textbf{Population Initialization:}  
The population \(\mathcal{P}\) comprises of \(m\) candidate solutions, defined as \(\mathcal{P} = \{ \mathbf{W}_{S_1}, \mathbf{W}_{S_2}, \ldots, \mathbf{W}_{S_m} \}\). The initial solution \(\mathcal{P}_0\) is set as the weight subset \(\mathbf{W}_{\text{sub}}\), and each subsequent solution \(\mathcal{P}_j\) is created by applying a mutation operation to \(\mathcal{P}_0\) as shown in Figure \ref{fig:AttentionBreaker}:

\begin{equation}
    \mathcal{P}_j \gets \texttt{Mutate}(\mathcal{P}_0).
\end{equation}

\noindent\textbf{Mutation Operation:}  
In the mutation process, each solution \(\mathcal{P}_j\) is assigned a unique mutation rate \(p_j\), randomly chosen per solution such that \(p_j \leq \mu\), the mutation rate. \(p_j\) then governs the probability of removal for each weight in \(\mathcal{P}_j\). Formally, for each weight \(w_{j_i} \in \mathcal{P}_j\), the mutated weight \(w_{j_i}'\) is defined as:

\begin{equation}
    w_{j_i}' = 
\begin{cases} 
\emptyset, & \text{with probability } p_j, \\
w_{j_i}, & \text{otherwise}.
\end{cases}
\end{equation}

In this context, ``mutation'' refers specifically to reducing the solution set’s cardinality rather than its traditional biological connotation. This reduction-driven mutation is designed to identify compact subsets that either preserve or improve model loss relative to the original subset. By systematically exploring minimal yet impactful configurations, this mutation strategy facilitates efficient traversal of the solution subspace and enhances the identification of subsets that sustain the desired adversarial effect on model performance.


\noindent\textbf{Elitism and Selection:}
At each iteration (\(t\)) of the optimization procedure, all solutions in \(\mathcal{P}\) are sorted by their fitness values:
\begin{equation}
    \mathcal{P}_{best} = \mathbf{W}_{S_{best}} = \arg \min_{\mathcal{P}_{j} \in \mathcal{P}} f(\mathcal{P}_{j})
\end{equation}
where \(\mathbf{W}_{S_{best}}\) or $\mathcal{P}_{best}$ represents the solution with the highest fitness (\textit{i.e.}, most efficient in degrading model performance). Elitism dictates that $\mathcal{P}_{best}$ is carried forward unchanged into the next generation, ensuring the highest-quality solution is preserved.
For the remaining solutions, a tournament selection strategy is employed to choose pairs of solutions as parents for crossover. For selecting each parent, this involves choosing two random solutions \(\mathcal{P}_{a}, \mathcal{P}_{b} \in \mathcal{P}\), and selecting the one with better fitness:
\begin{equation}
    \mathcal{P}_{p} = \arg \min \{ f(\mathcal{P}_{a}), f(\mathcal{P}_{ b}) \}.
\end{equation}
This strategy is chosen over a random strategy due to its capability to balance exploration and exploitation in the search space, thereby enhancing convergence toward an optimal solution.




\begin{algorithm}[t]
\caption{Weight Set Optimization - GenBFA}
\begin{algorithmic}[1]
\Require $\mathbf{W}$, Weight subset $\mathbf{W}_{sub}$, Target loss threshold $\mathcal{L}_{th}$, population size $m$, max generations $g$, mutation/reduction rate $\mu$, no-improvement threshold $\mathcal{N}$
\State \text{Initialize solution space }$\mathbf{SS} = [\text{ }], \text{ solution set }\mathcal{P} = [\text{ }], \text{ fitness scores } \mathbf{F} = [\text{ }]$, no-improvement counter $c = 0$ 
\State $l,\mathbf{SS} \gets \mathbf{W}_{sub}[0], \mathbf{W}_{sub}[1]$
\State $\mathcal{P} \gets \texttt{InitializePopulation}(\mathbf{SS},\mu)$
\For{$t = 1, 2, \dots, g$}
\For{$j = 1,2,...,m$}
\State $\mathbf{W}^{(l)} = \texttt{GetParams}(\mathbf{W},l)$
\State $I^{(l)}_{hybrid} \gets \mathcal{P}_{j}$
    \State $\mathcal{L}_j \leftarrow \texttt{BFLIP}
    ({\mathbf{W}^{(l)}, pos, I^{(l)}_{hybrid}})$
    \State $f \leftarrow \operatorname{sgn}\left(\mathcal{L}_j - \mathcal{L}_{th}\right)*\frac{\mathcal{L}_{S_j}}{\texttt{cardinality}(\mathbf{W}^{(l)})}$
    
\State \text{Push} $f$ \text{to} $\mathbf{F}$
\EndFor
    \State $\mathcal{P} \gets \texttt{SortSolutions}({\mathcal{P},\mathbf{F}})$ 
    \State $\mathcal{P}_{best}^{old} \gets$ {$\mathcal{P}_{best}$} 
    \State $\mathcal{P}_{best} \gets$ \texttt{SelectBestSolution}({$\mathcal{P}$}) 
    \State \texttt{CheckTerminate}($\mathcal{P}_{best},\mathcal{P}_{best}^{old},c,\mathcal{N}$)
    \State $\mathcal{P}_{new} \gets [\mathcal{P}_{best},\texttt{Mutate}(\mathcal{P}_{best})]$ 
    \While{$\texttt{len}(\mathcal{P}_{new}) < m$}
        \State $\mathcal{P}_{p1}, \mathcal{P}_{p2} \gets$ \texttt{TournamentSelection}($\mathcal{P},\mathbf{F}$)
        \State $\mathcal{P}_{o1}, \mathcal{P}_{o2} \gets$ \texttt{Crossover}($\mathcal{P}_{best},\mathcal{P}_{p1}, \mathcal{P}_{p2}$)
        \State $\mathcal{P}_{o1}, \mathcal{P}_{o2} \gets$ \texttt{Mutate}{($\mathcal{P}_{o1}, \mu$)},\texttt{Mutate}{($\mathcal{P}_{o2}, \mu$)}  
        \State Push $\mathcal{P}_{o1}$ and $\mathcal{P}_{o2}$ to $\mathcal{P}_{new}$
    \EndWhile
    \State $\mathcal{P} \gets \mathcal{P}_{new}$ \Comment{Population for next generation}
\EndFor
\Ensure $\mathcal{P}_{best}$ \Comment{Return best solution}
\end{algorithmic}
\label{algo:genetic}
\end{algorithm}
\noindent\textbf{Crossover Operation:}
To generate new solutions, two-parent solutions, \( \mathcal{P}_{p1} \) and \( \mathcal{P}_{p2} \), are selected, followed by a crossover operation executed with probability \( p_c \). During crossover, the current best solution \( \mathcal{P}_{\text{best}} \) is integrated to propagate its beneficial traits forward. Two offspring, \( \mathcal{P}_{o1} \) and \( \mathcal{P}_{o2} \), are produced by combining \( \mathcal{P}_{\text{best}} \) with each parent \( \mathcal{P}_{p1} \) and \( \mathcal{P}_{p2} \) independently. 

For offspring \(\mathcal{P}_{o1} \), each gene \( g_i \) is chosen stochastically from either \(\mathcal{P}_{p1} \) or \( \mathcal{P}_{\text{best}} \), represented as:
\begin{equation}
    \mathcal{P}_{o1} = \{ g_{\mathcal{P}_{p1}}^i \text{ or } g_{\mathcal{P}_{\text{best}}}^i \} \quad \forall i
\end{equation}

Similarly, offspring \( \mathcal{P}_{o2} \) is created by selecting each gene \( g_i \) stochastically from \( \mathcal{P}_{p2} \) or \( \mathcal{P}_{\text{best}} \):
\begin{equation}
    \mathcal{P}_{o2} = \{ g_{\mathcal{P}_{p2}}^i \text{ or } g_{\mathcal{P}_{\text{best}}}^i \} \quad \forall i
\end{equation}

If crossover does not occur (i.e., with probability \( 1 - p_c \)), the offspring remain identical to the parents, such that \( \mathcal{P}_{o1} = \mathcal{P}_{p1} \) and \( \mathcal{P}_{o2} = \mathcal{P}_{p2} \).

\noindent \textbf{Convergence and Termination:} The process iterates over generations indexed by \( t = 1, 2, \dots, g \), where \( g \) is the predefined maximum number of iterations. The algorithm terminates when either the iteration limit \( g \) is reached (\( t = g \)) or a predefined no-improvement limit \( \mathcal{N} \) is met, ensuring computational efficiency. The final output is the weight subset \( \mathbf{W}_{opt} \subset \mathbf{W}_{sub} \), where

\[
\mathbf{W}_{opt} = \arg \min_{\mathcal{P}_j \in \mathcal{P}} f(\mathcal{P}_j),
\]

represents the critical weights required to achieve the desired model loss degradation. 

The overall optimization process is outlined in Algorithm \ref{algo:genetic}, which begins with model weights \(\mathbf{W}\), a target loss threshold \(\mathcal{L}_{th}\), and parameters defining population characteristics: population size \(m\), maximum iterations \(g\), crossover probability \(p_c\), and mutation rate \(\mu\).
The algorithm initializes solution space \(\mathbf{SS}\), population \(\mathcal{P}\), and fitness scores \(\mathbf{F}\) (\textit{line 1}). The attacked layer \(l\) and initial solution space are generated from \(\mathbf{W}_{sub}\) (\textit{line 2}). The initial solution \(\mathcal{P}_0\) (=$\mathbf{W}_{sub}$) is used to create a population by applying \(m-1\) mutations, resulting in \(\mathcal{P}\) (\textit{line 3}).

At each iteration \(t\), solutions are evaluated using the \texttt{BFLIP} function, which computes model loss \(\mathcal{L}_j\), and the fitness score is determined based on \(\mathcal{L}_j\), \(\mathcal{L}_{th}\), and the \texttt{cardinality} of \(\mathbf{W}^{(l)}\) using Equation \ref{eq:loss_fn}(\textit{lines 6-10}). Solutions are ranked by fitness (\textit{line 12}).

The current best solution is tallied against the previous best solution for capturing improvement and checking for termination using \texttt{CheckTerminate} function (\textit{lines 13-15}).
To ensure elitism, the best solution \(\mathcal{P}_{best}\) is carried over to the next generation \(\mathcal{P}_{new}\) (\textit{line 16}). The second solution is created by mutating \(\mathcal{P}_{best}\), and the remaining population is filled by generating offspring through crossover and mutation (\textit{lines 17-20}). Crossover selects parent pairs via tournament selection (\textit{line 18}), generates offspring (\textit{line 19}), and applies mutation (\textit{line 20}). These offspring are added to \(\mathcal{P}_{new}\) (\textit{line 21}).
The new population \(\mathcal{P}_{new}\) replaces \(\mathcal{P}\) (\textit{line 23}), and the process repeats until either \(g\) generations or the no-improvement threshold \(\mathcal{N}\) is reached. The algorithm outputs \(\mathcal{P}_{best}\) as the optimal weight subset.
\section{Experimental results}\label{sec:results}
This section presents experimental results showcasing the effectiveness of AttentionBreaker across different model precisions, types, and tasks.

\subsection{Experimental Setup}



For our experiments, we utilized a range of models, including the open-source LLaMA3-8B-Instruct~\cite{touvron2023llama, dubey2024llama}, Phi-3-mini-128k-Instruct~\cite{abdin2024phi} and BitNet ~\cite{ma2024era}. The evaluation of LLMs was conducted using standard benchmarks like MMLU~\cite{mmlu} and the Language Model (LM) Evaluation Harness\cite{eval-harness}, both of which test the models' ability to reason and generalize across a variety of tasks. The tasks we used from the LM Harness used for LLM performance analysis are ``AddSub'', ``ArcE'', ``ArcC'', ``BoolQ'', ``HellaSwag'', ``MathQA'', ``PIQA'', ``SocialIQA'', ``SVAMP'', ``OpenBookQA'' and ``GSM8k'' (indexed from 0 to 10). 
As multi-modal models continue to gain prominence, we extended our evaluation to include a Vision-Language Model (VLM), LLaVA1.6~\cite{liu2023improved}. The performance of this VLM was assessed using the VQAv2~\cite{nguyen2023openvivqa} and TextVQA~\cite{singh2019towards} benchmarks, which are widely recognized metrics for quantifying multi-modal reasoning capabilities.
To demonstrate the versatility of AttenBreaker, we test it against several quantization formats, including the standard 8-bit integer (INT8), normalized float 4 (NF4), and an advanced 1.58-bit precision format implemented using the BitNet framework~\cite{ma2024era}. Please note that all these models are weight-only quantized, and therefore, we will use notations W8, W4, and W1.58, respectively. These lower-bit precision formats lower computational requirements and memory usage while maintaining their performance in metrics such as perplexity for LLMs and classification accuracy for VLMs. In our experimental setup, only the attention and the Multi-Layer Perceptron (MLP) layers of the models are quantized. In this study, attacks are executed only for these layers, and their effects are analyzed.

The experimental setup utilized 4x NVIDIA A100 GPUs, each equipped with 80GB of memory, and operated under CUDA version 12.2.


\begin{figure*}[t!]
  \centering
    \includegraphics[width=0.945\linewidth]
{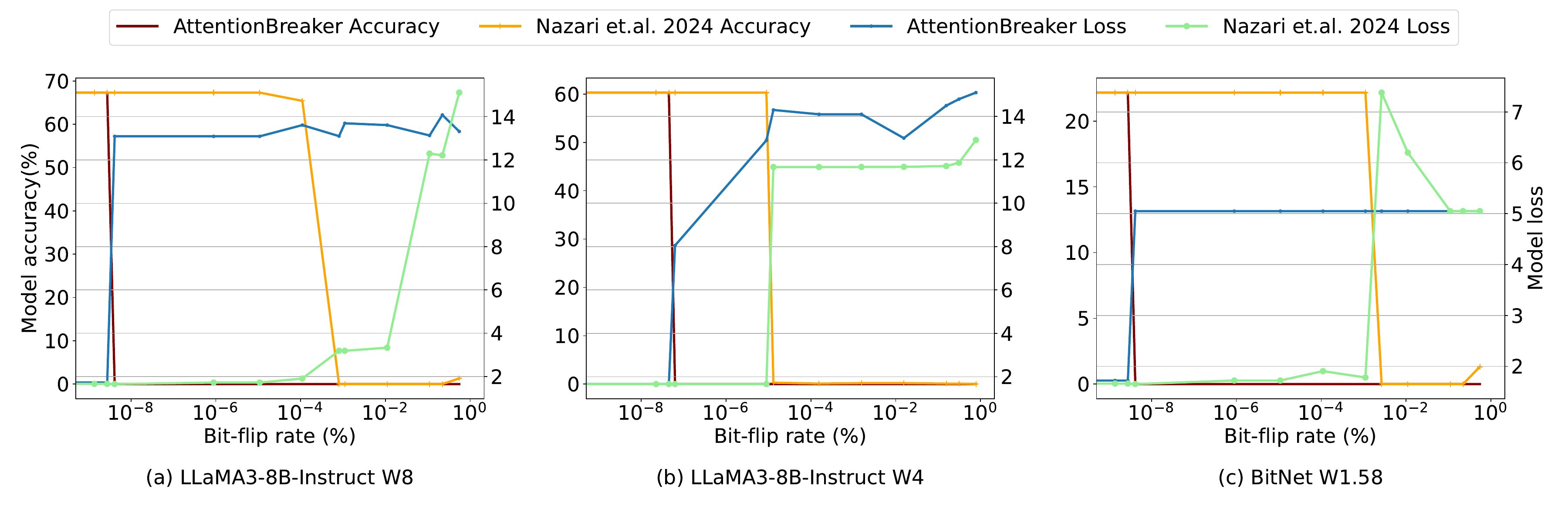}
\vspace{-2mm}
  \caption{Comparison of BFA attack using \cite{nazari2024forget} and AttentionBreaker on LLaMA3-8B-Instruct W8, W4, and W1.58 models. Metrics are calculated on the ``Astronomy'' task of the MMLU benchmark.}
  \label{fig:wsweep}
  \vspace{-2mm}
\end{figure*}


To evaluate the model performance, we calculate the perplexity score~\cite{hu2024can} and accuracy on a variety of benchmarks.
Perplexity is a measure of how well a probability model predicts a sample. It is calculated as the exponential of the average negative log-likelihood of a sequence as shown in Equation \ref{eq:perplexity}.

\begin{equation}
\text{Perplexity} = \exp\left(-\frac{1}{N} \sum_{i=1}^{N} \log P(x_i)\right) 
\label{eq:perplexity}
\end{equation}

where $N$ is the number of words, and $P(x_i)$ is the probability assigned to the $i$-th word by the model. A lower perplexity indicates better performance. Perplexity is calculated on the Wikitext~\cite{merity2016pointer} dataset.

Accuracy on benchmarks is another critical metric, especially for tasks like classification~\cite{gupta2024changing}. 
It is the ratio of correctly predicted instances to the total instances:
\begin{equation}
\text{Accuracy} = \frac{1}{N} \sum_{i=1}^{N} \mathbb{I}(\hat{y}_i = y_i)
\label{eq:acc}
\end{equation}
Where \( N \) is the total number of tasks. \( \hat{y}_i \) is the predicted label for the \( i \)-th task. \( y_i \) indicates the true label for the \( i \)-th task. \( \mathbb{I} \) is the indicator function that returns 1 if the predicted label matches the true label and 0 otherwise.

\subsection{Results and Analysis}
This section presents the experimental results demonstrating the efficacy of the AttentionBreaker framework. Primary outcomes are highlighted here, supported by intermediate analyses in Section \ref{sec:intermediate}.
\subsubsection{AttentionBreaker on 8-bit Quantized LLMs}
Here, we present the efficacy of the proposed AttentionBreaker framework on 8-bit weight quantized (W8) LLMs. Table \ref{table:main} compares the effects of existing Bit-Flip Attack \cite{nazari2024forget} and AttentionBreaker on 8-bit versions of LLM across several benchmarks.
For each model, the precision, number of parameters, and bit-flip Count required to disrupt performance are listed in the Table. 
Benchmark results show that AttentionBreaker consistently reduces accuracy to zero across datasets, proving more efficient than existing BFA in degrading model performance with fewer bit-flips. The results highlight the impact of bit-flips on key model performance metrics, including perplexity, MMLU, and LM harness benchmarks.

For instance, LLaMA3-8B-Instruct W8 suffers accuracy drop to $0$\% on MMLU tasks from the original accuracy of $67.3$\%  by merely  \textbf{3 bit-flips} ($\mathbf{4.129 \times 10^{-9}}$\% bit-flip rate) as captured in Figure 6(a). At the same time, the perplexity measure increases from $12.14$ to \(4.72\times10^5\) post-attack. In comparison, \cite{nazari2024forget} require \textbf{$\textbf{10}^\textbf{7}\times$ ($\approx \textbf{0.1}$\% bit-flip rate)} more bit-flips to accomplish a similar level of model degradation. A similar trend is observed for Phi-3-mini-128k-Instruct W8 models as well.

\subsubsection{AttentionBreaker on 4-bit quantized LLMs}
In this section, we present the performance of AttentionBreaker on 4-bit weight quantized (W4) LLMs. As shown in Figure 6(b), the attack on LLaMA3-8B W4 demonstrates that $\mathbf{7.8125 \times 10^{-8}}$\% or \textbf{28 bit-flips} results in a perplexity increase from $12.60$ to $3.8$$\times10^4$, with the MMLU score dropping from $60.3$\% to $0$. A similar trend is observed for Phi-3-mini-128k-Instruct W4 models as well. \textcolor{black}{\textit{This showcases the efficiency of the proposed AttentionBreaker framework.}} 


\subsubsection{AttentionBreaker on low-precision LLM}
We further evaluate the effectiveness of our approach for ultra-low precision models, such as BitNet W1.58\cite{ma2024era}. BitNet W1.58 represents a 1.58-bit variant of LLMs where each model parameter (or weight) is constrained to ternary values \(\{-1, 0, 1\}\). 
Given their low precision, these models are intuitively expected to be inherently resistant, as a single bit-flip induced error has limited impact. Employing our approach, we observe that $45811 \text{ or } 5\times 10^{-4}$\%  bit-flip rate is required to degrade the model performance to near $0$\% effectively as demonstrated in Figure 6(c). In comparison, \cite{nazari2024forget} require $10^5\times$ ($\approx 1$\% bit-flip rate) more bit-flips to accomplish a similar level of model degradation. \emph{This demonstrates the efficacy of AttentionBreaker in efficiently attacking low-precision models as well. }

\begin{table*}[h!]
\caption{AttentionBreaker attack results on various models and datasets.}
\renewcommand{\arraystretch}{1.3}
\resizebox{1\textwidth}{!}{\begin{tabular}{|c|c|c|c|c|c|cccllllcc|}

\hline
\rowcolor[HTML]{C0C0C0} 
\cellcolor[HTML]{C0C0C0}                                 & \cellcolor[HTML]{C0C0C0}                                                                                & \cellcolor[HTML]{C0C0C0}                                     & \cellcolor[HTML]{C0C0C0}                                                                                           & \cellcolor[HTML]{C0C0C0}                                                                                                  & \cellcolor[HTML]{C0C0C0}                                                                                                        & \multicolumn{9}{c|}{\cellcolor[HTML]{C0C0C0}\textbf{Benchmarks (Accuracy in \%) (Before attack / After Attack)}}                                                                                                                                                                                                               \\ \cline{7-15} 
\rowcolor[HTML]{C0C0C0} 
\multirow{-2}{*}{\cellcolor[HTML]{C0C0C0}\textbf{Model}} & \multirow{-2}{*}{\cellcolor[HTML]{C0C0C0}\textbf{\begin{tabular}[c]{@{}c@{}}Model\\ type\end{tabular}}} & \multirow{-2}{*}{\cellcolor[HTML]{C0C0C0}\textbf{Precision}} & \multirow{-2}{*}{\cellcolor[HTML]{C0C0C0}\textbf{\begin{tabular}[c]{@{}c@{}}Number of \\ Parameters\end{tabular}}} & \multirow{-2}{*}{\cellcolor[HTML]{C0C0C0}\textbf{\begin{tabular}[c]{@{}c@{}} BFA\cite{nazari2024forget}\\ (\# of bit-flips)\end{tabular}}} & \multirow{-2}{*}{\cellcolor[HTML]{C0C0C0}\textbf{\begin{tabular}[c]{@{}c@{}}AttentionBreaker\\ (\# of bit-flips)\end{tabular}}} & \multicolumn{1}{c|}{\cellcolor[HTML]{C0C0C0}\textbf{WikiText-2}} & \multicolumn{1}{c|}{\cellcolor[HTML]{C0C0C0}\textbf{MMLU}} & \multicolumn{5}{c|}{\cellcolor[HTML]{C0C0C0}\textbf{LM Harness}} & \multicolumn{1}{l|}{\cellcolor[HTML]{C0C0C0}\textbf{VQAv2}} & \multicolumn{1}{l|}{\cellcolor[HTML]{C0C0C0}\textbf{TextVQA}} \\ \hline \hline
                                                         &                                                                                                         & 4-bit NF4                                                    &                                                                                                                    & 3.4$\times10^8$                                                                                                                     & 28                                                                                                                              & \multicolumn{1}{c|}{12.60/3.8$\times10^4$}                                 & \multicolumn{1}{c|}{60.3/0}                                & \multicolumn{5}{c|}{61.2/0}                                      & \multicolumn{1}{c|}{NA}                                     & NA                                                            \\ \cline{3-3} \cline{5-15} 
\multirow{-2}{*}{LLaMA3-8B-Instruct}                                & \multirow{-2}{*}{LLM}                                                                                   & INT8                                                         & \multirow{-2}{*}{8.03B}                                                                                            & 9.6$\times10^8$                                                                                                                     & 3                                                                                                                               & \multicolumn{1}{c|}{12.14/4.7$\times10^5$}                                 & \multicolumn{1}{c|}{67.3/0}                                & \multicolumn{5}{c|}{72.6/0}                                      & \multicolumn{1}{c|}{NA}                                     & NA                                                            \\ \hline
                                                         &                                                                                                         & 4-bit NF4                                                    &                                                                                                                    & 3.2$\times10^8$                                                                                                                     & 53                                                                                                                              & \multicolumn{1}{c|}{NA}                                          & \multicolumn{1}{c|}{NA}                                    & \multicolumn{5}{c|}{NA}                                          & \multicolumn{1}{c|}{75.6/0}                                 & 60.5/0                                                        \\ \cline{3-3} \cline{5-15} 
\multirow{-2}{*}{LLaVA1.6-Mistral-7B}                      & \multirow{-2}{*}{VLM}                                                                                   & INT8                                                         & \multirow{-2}{*}{7.57B}                                                                                            & 8.3$\times10^8$                                                                                                                     & 15                                                                                                                              & \multicolumn{1}{c|}{NA}                                          & \multicolumn{1}{c|}{NA}                                    & \multicolumn{5}{c|}{NA}                                          & \multicolumn{1}{c|}{81.4/0}                                 & 64.2/0                                                        \\ \hline
BitNet-large                                             & 1.58-bit LLM                                                                                            & 1.58-bit                                                     & 729M                                                                                                               & 2$\times10^5$                                                                                                                       & 45811                                                                                                                           & \multicolumn{1}{c|}{17.21/5.5$\times10^4$}                                 & \multicolumn{1}{c|}{22.1/0}                                & \multicolumn{5}{c|}{23.2/0}                                      & \multicolumn{1}{c|}{NA}                                     & NA                                                            \\ \hline
                                                         &                                                                                                         & 4-bit NF4                                                    &                                                                                                                    & 1.6$\times10^8$                                                                                                                     & 9                                                                                                                               & \multicolumn{1}{c|}{11.94/3.6$\times10^4$}                                 & \multicolumn{1}{c|}{66.1/0}                                & \multicolumn{5}{c|}{71.7/0}                                      & \multicolumn{1}{c|}{NA}                                     & NA                                                            \\ \cline{3-3} \cline{5-15} 
\multirow{-2}{*}{Phi-3-mini-128k-Instruct}                        & \multirow{-2}{*}{LLM}                                                                                   & INT8                                                         & \multirow{-2}{*}{3.82B}                                                                                            & 4.5$\times10^8$                                                                                                                     & 4                                                                                                                               & \multicolumn{1}{c|}{11.82/3.7$\times10^4$}                                 & \multicolumn{1}{c|}{69.7/0}                                & \multicolumn{5}{c|}{72.3/0}                                      & \multicolumn{1}{c|}{NA}                                     & NA                                                            \\ \hline
\end{tabular}}
\label{table:main}
\end{table*}

\subsubsection{AttentionBreaker on VLM}
We demonstrate the efficacy of AttentionBreaker on multimodal models such as VLMs, which integrate a vision encoder with a language model to enable multimodal tasks, including image comprehension, alongside text-based tasks. When applying our attack methodology to the LLaVA1.6-7B W8 model, which comprises 7.57 billion parameters, we observe that a minimal bit-flip rate of $\mathbf{2.394 \times 10^{-8}}$\%—equivalent to only \textbf{15 bit-flips}—causes a substantial degradation in performance. Specifically, the VQAv2 and TextVQA scores, initially at $81.4\%$ and $64.2\%$, respectively, drop to $0\%$ following the attack. In comparison, \cite{nazari2024forget} require ($\mathbf{10^7\approx 0.12}$\% bit-flip rate) more bit-flips to accomplish a similar level of model degradation. \emph{This result underscores the efficacy of the AttentionBreaker framework in exploiting latent vulnerabilities within multimodal LLMs as well.}


\subsection{Intermediate Results}\label{sec:intermediate}
This section outlines intermediate results that validate assumptions and inform the main findings in Section \ref{sec:results}. 
\subsubsection{Assessing Layer Sensitivity}
\label{subsec:Layer_sensitivity_analysis}


\begin{figure*}[h!]
  \centering
  \begin{subfigure}{0.24\linewidth}
    \includegraphics[width=1.0\linewidth]
{figures/llama_3_8b_instruct_int8_layer_sens_1_v2.pdf}
    \caption{LLaMA3-8B W8.}
    \label{fig:lsensLLaMA3-8B8}
  \end{subfigure}
  \begin{subfigure}{0.24\linewidth}
    \includegraphics[width=1\linewidth]
{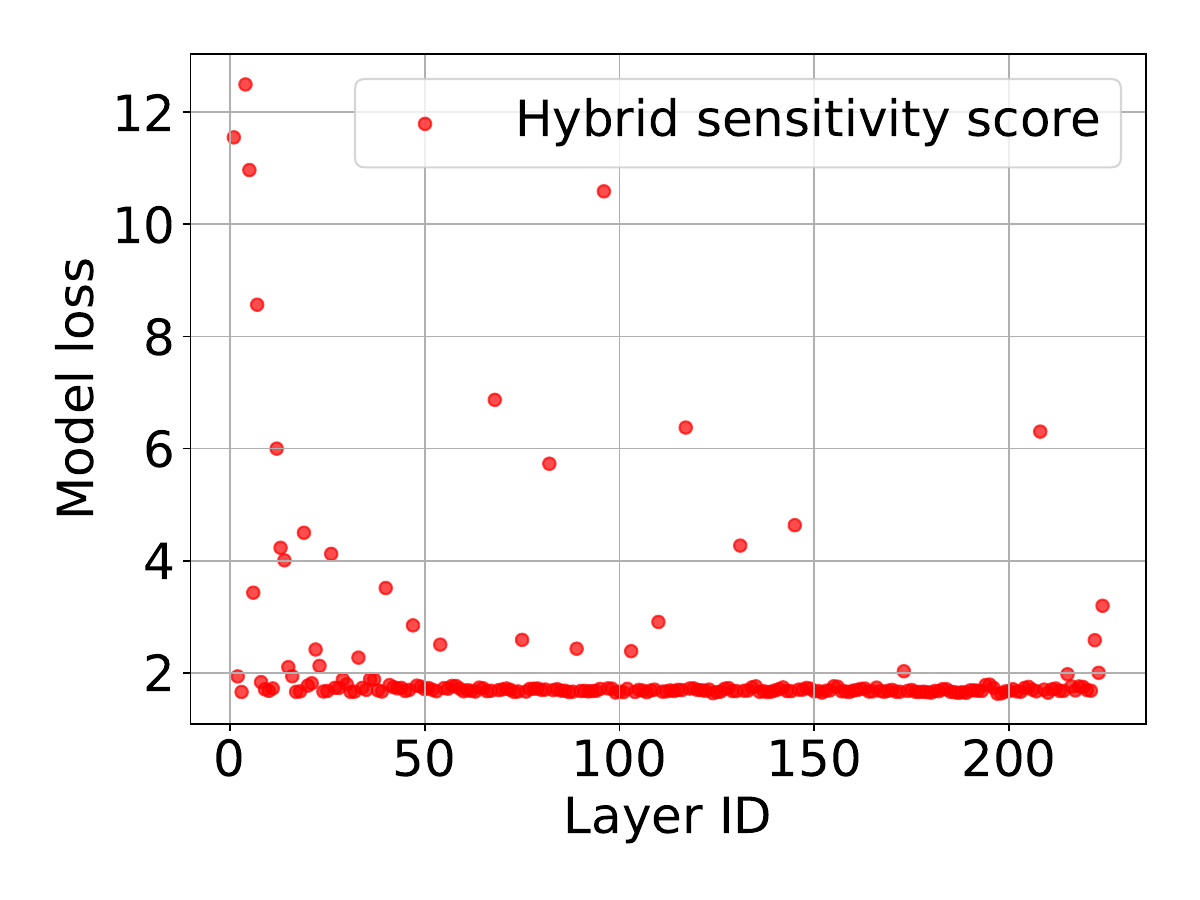}
    \caption{LLaMA3-8B W4.}
    \label{fig:lsensLLaMA3-8B4}
  \end{subfigure}
  \begin{subfigure}{0.24\linewidth}
    \includegraphics[width=1\linewidth]
{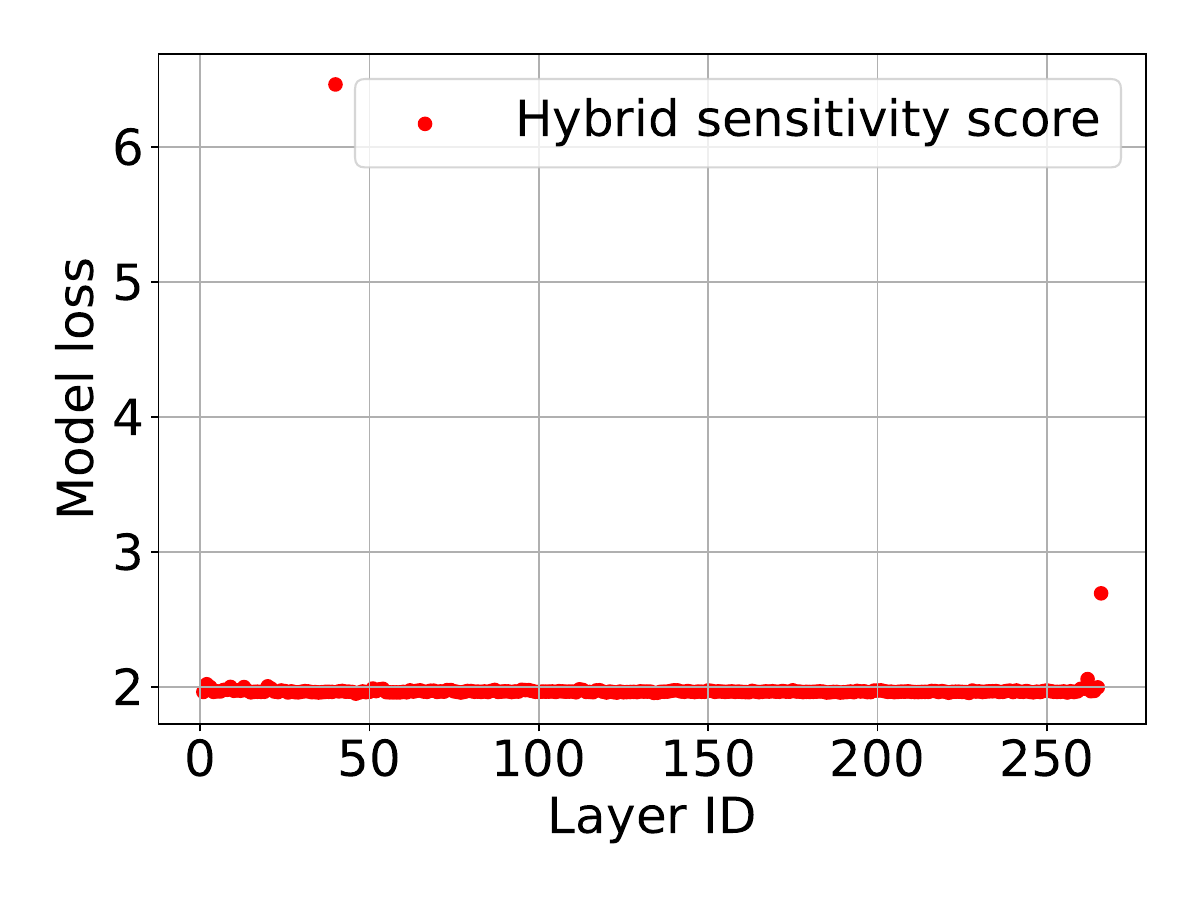}
    \caption{BitNet W1.58.}
    \label{fig:lsensbitnet}
  \end{subfigure}
  \begin{subfigure}{0.24\linewidth}
    \includegraphics[width=1\linewidth]
{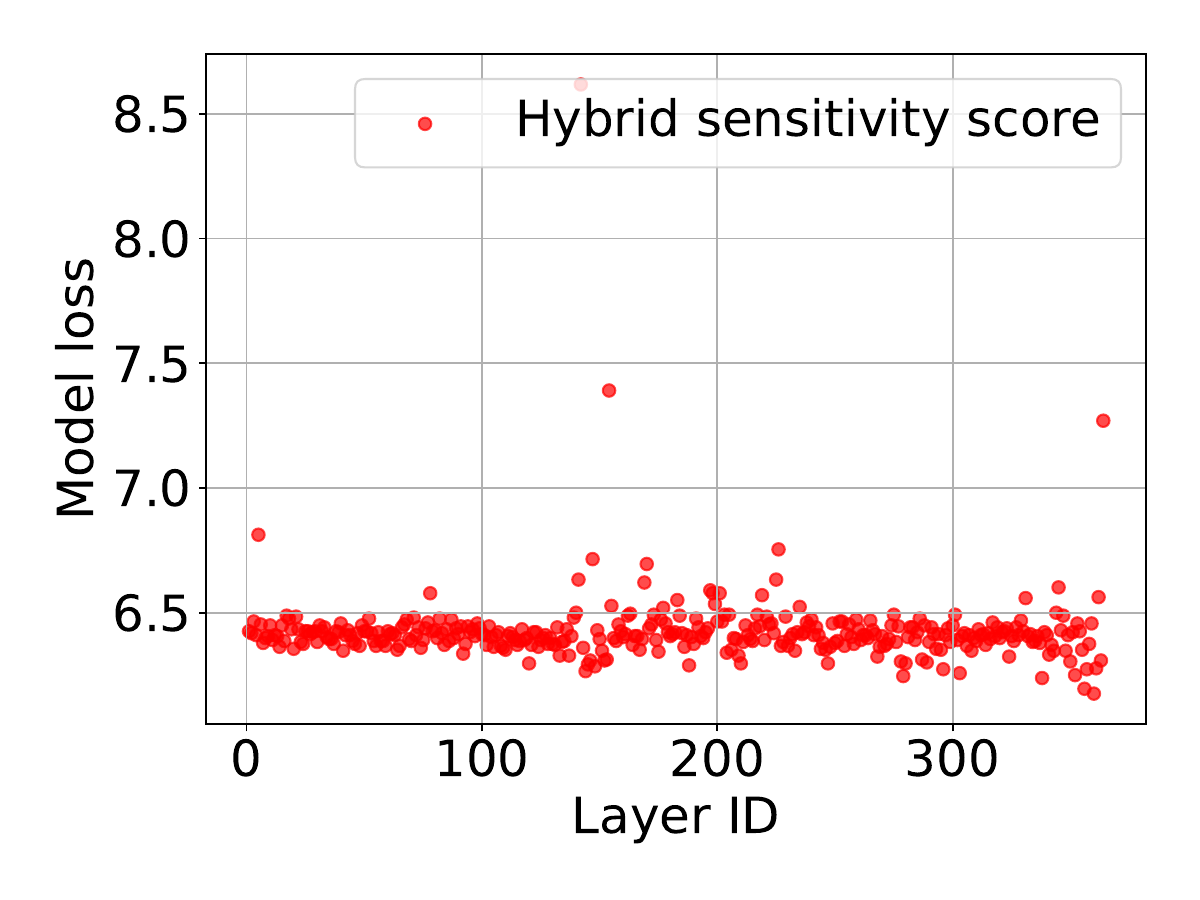}
    \caption{LLaVA1.6-7B W8.}
    \label{fig:lsensllava}
  \end{subfigure}
  \caption{Layer sensitivity analysis for LLMs and VLMs with varying quantization formats. Sensitivity scores and model loss are calculated using the `Astronomy' task of the MMLU benchmark.}
  \label{fig:layer_sens}
\end{figure*}
Based on the layer sensitivity analysis described \ref{sec:layer_sensitivity}, we obtain layer sensitivity metrics. Layers are subsequently ranked according to this sensitivity measure, from the most sensitive—those yielding the highest impact on loss as calculated through cross-entropy~\cite{zhang2018generalized}—to the least sensitive. 
Figure \ref{fig:layer_sens} presents the sensitivity distribution across layers for the LLaMA3-8B-Instruct  W8 and W4, along with BitNet and LLaVA1.6-7B models, illustrating the effects of bit-flip perturbations on different layers. Sensitivity variations are quantified through model loss responses to targeted perturbations at quantized layers, indicated by Layer ID. For LLaMA3-8B-Instruct, bit-flips on INT8-quantized layers reveal high sensitivity in initial layers, as depicted in Figure \ref{fig:lsensLLaMA3-8B8}. When LLaMA3-8B-Instruct is quantized to NF4, the model loss remains elevated, signifying high sensitivity in Figure \ref{fig:lsensLLaMA3-8B4}. Figure \ref{fig:lsensbitnet} describes the sensitivity of BitNet, which, with its more aggressive quantization, exhibits greater resilience, reflected by fewer layers furnishing high model loss. In the case of LLaVA1.6-7B at INT8, high sensitivity to parameter perturbation is observed primarily in the language processing component of the model. The vision processing component, preceding the language component, has a less pronounced effect. The most sensitive layers are among the quantized layers at the beginning of the language component, as seen in Figure \ref{fig:lsensllava}. 
Based on this layer sensitivity metric, the layers are sorted, and top-$5$ layers are chosen for the subsequent experiments, thereby reducing the search space.
Notably, for LLaMA3-8B, selecting the top-5 most sensitive layers allowed for a reduction in the solution space from $\approx 8\times10^9$ to $\approx 2.1\times10^8$ parameters representing a reduction of $\approx 38\times$.

\subsubsection{Selecting Weights from the Most Sensitive Layer(s)}\label{sec:wight_selection_an}

In this section, we select a subset of weights from the top-$5$ layers obtained after layer sensitivity analysis as discussed in Section \ref{sec:subset_selection}.
\textcolor{black}{The sampled subsets of weights were evaluated against a loss threshold. The threshold was empirically set to 7 for most models, with adjustments made for certain models to match their baseline loss levels on the specific task. This threshold selection aimed to balance solution quality with computational efficiency: a higher threshold would impose greater constraints during optimization, potentially increasing convergence time but yielding solutions with higher loss, which could improve BFA effectiveness. Conversely, a lower threshold would ensure faster convergence but might result in suboptimal solutions.}  
Layers that met this threshold were then ranked in ascending order by the number of bit-flips required, resulting in the smallest subset of weights that maintains the degraded model performance.
This results in a significant decrease in the solution size, achieving up to a $10^5 \times$ reduction, which allows the weight-set optimization phase (described in Section \ref{sec:genetic}) to operate on a much smaller subset of weights substantially enhancing both computational efficiency and scalability. For example, in the LLaMA3-8B Instruct W8 model, we selected the top-5 layers for the subset selection experiment, which together contained $\approx 2.1\times10^8$ parameters. After applying the subset selection technique, we reduced the weight subset to 5,872 parameters, representing a reduction factor of $\approx3.5 \times 10^4$. This analysis demonstrates the capability of the framework to streamline the optimization process.

\subsubsection{Obtaining the Most Critical Weight-set}

\begin{figure*}[h!]
  \centering
  \begin{subfigure}{0.32\linewidth}
    \includegraphics[width=1\linewidth]
{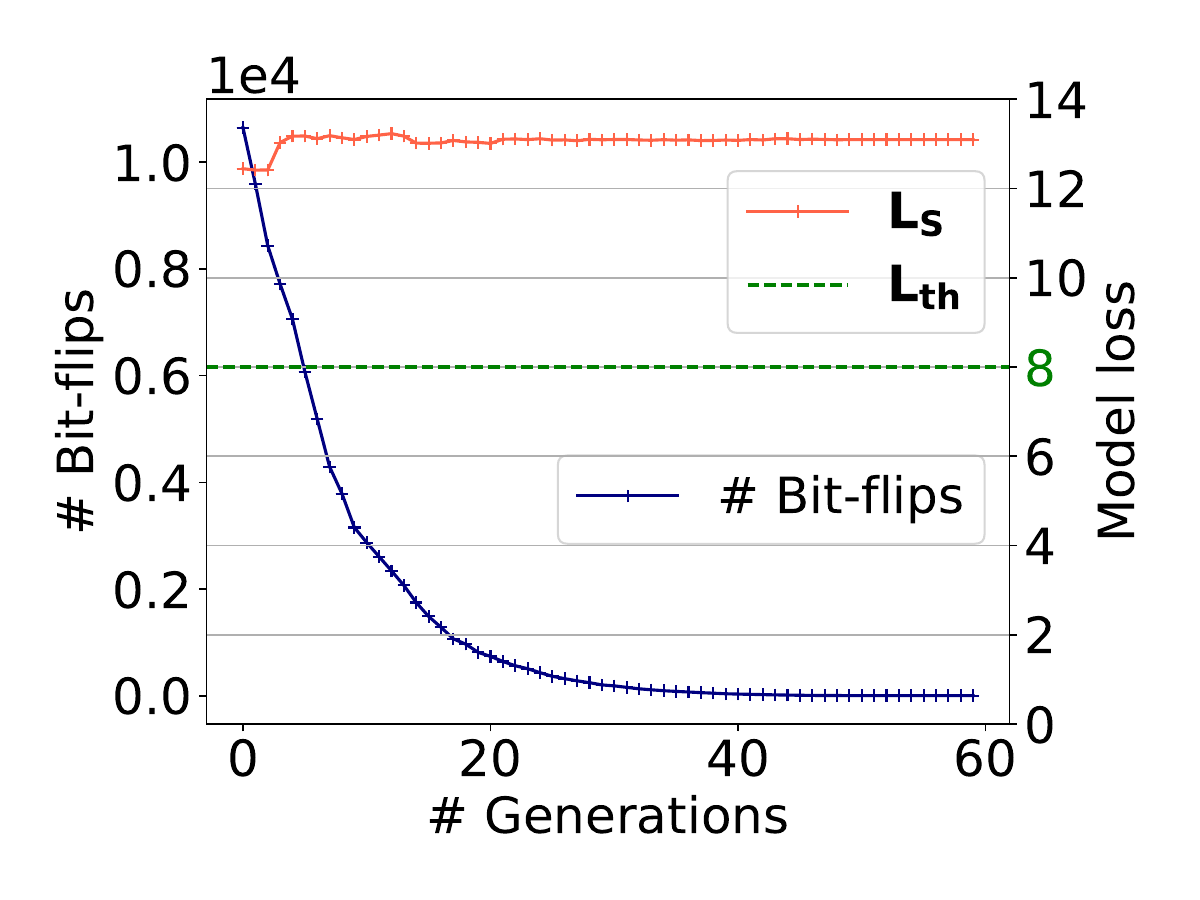}
    \caption{LLaMA3-8B-Instruct-Instruct W8.}
  \end{subfigure}
  \begin{subfigure}{0.32\linewidth}
    \includegraphics[width=1\linewidth]
{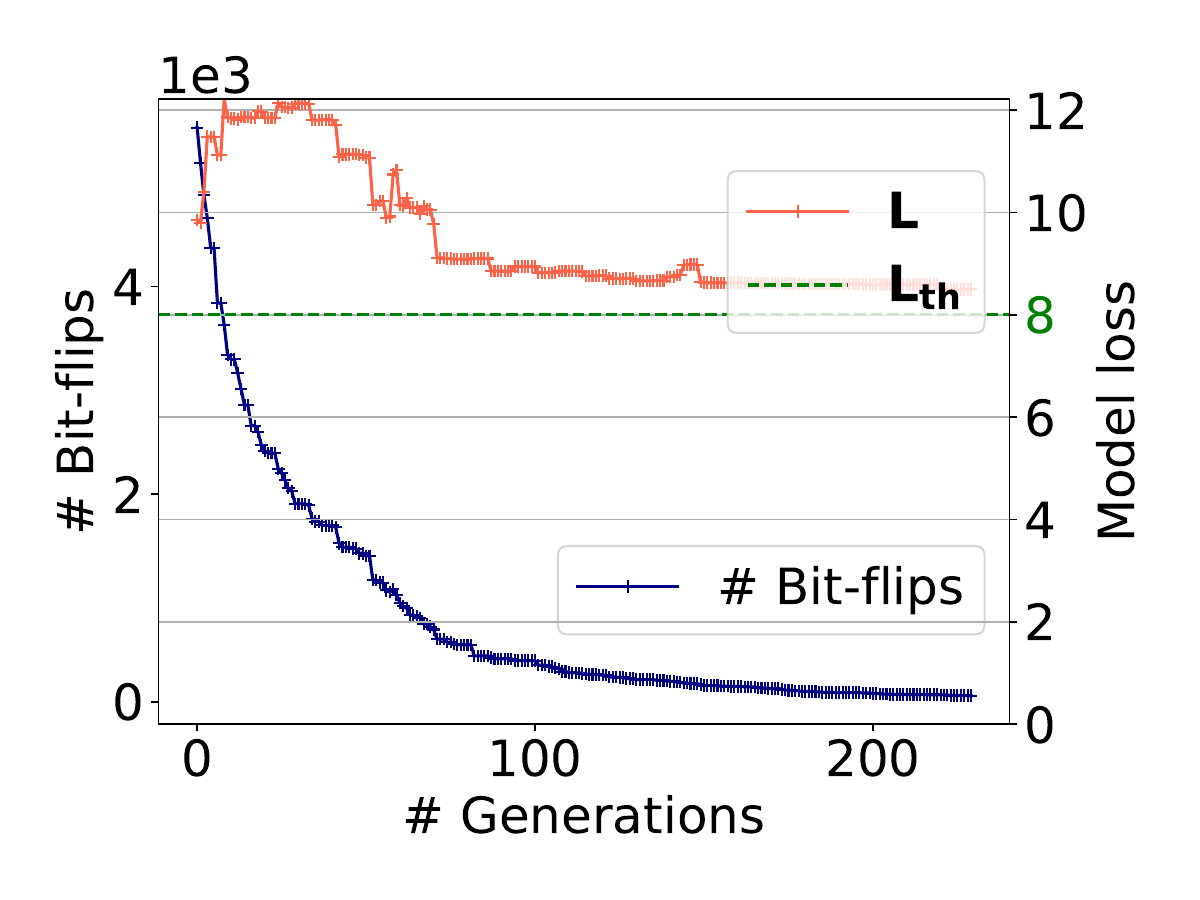}
    \caption{LLaMA3-8B-Instruct W4.}
    \label{fig:gensweep}
  \end{subfigure}
  \begin{subfigure}{0.32\linewidth}
    \includegraphics[width=1\linewidth]
{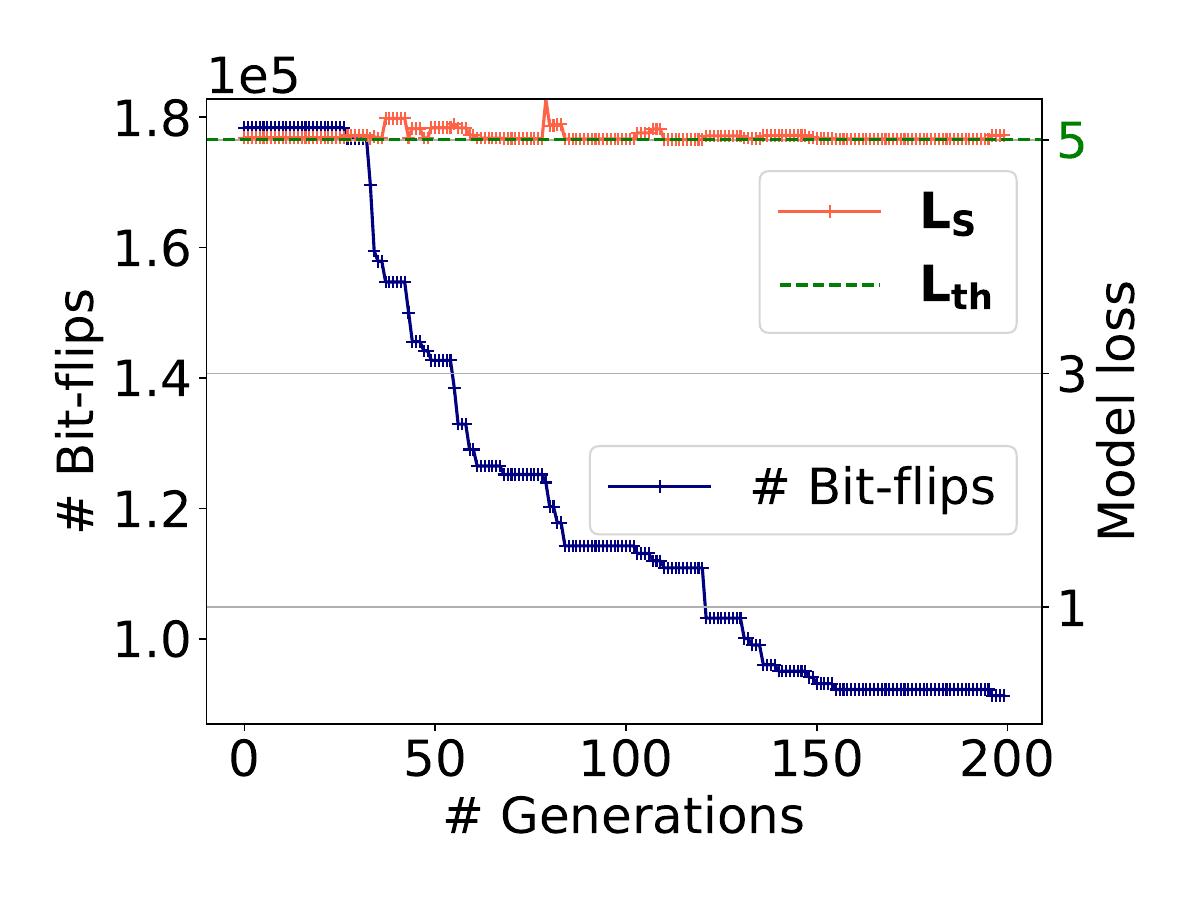}
    \caption{BitNet W1.58.}
  \end{subfigure}
  \caption{Genetic optimization for critical parameter identification. Loss is calculated on the `Astronomy' task on MMLU.}
  \label{fig:geneticopt}
\end{figure*}

The GenBFA algorithm, detailed in Section \ref{sec:genetic}, refines the weight subset generated by the previous step. 
This iterative optimization process is captured in Figure \ref{fig:geneticopt}, which shows an exponential reduction in weight subset size while maintaining loss at or higher than the loss threshold.
For the LLaMA3-8B Instruct W8 model, the initial subset size is reduced by up to \(2 \times 10^3\) (see Figure \ref{fig:lsensLLaMA3-8B8}), with only \textbf{3} weights needed to execute the AttentionBreaker attack at a level of efficacy comparable to perturbing the original set of 5872 weights. Results for 8-bit, 4-bit LLaMA, and BitNet models, shown in Figure 6, underscore the optimization approach’s effectiveness, demonstrating its adaptability across models with varying precision levels.

\subsection{Adaptability and Robustness}
In this section, we examine the adaptability of AttentionBreaker in constrained environments, along with its robustness and the transferability of its attack effects.
\subsubsection{Gradient-free AttentionBreaker}\label{sec:gradfree}
In this experiment, we investigate the effectiveness of AttentionBreaker in a gradient-free context, where gradient information is inaccessible at all stages. Here, the importance scores are computed solely based on weight magnitudes by setting \( \alpha = 0.0 \). Our results demonstrate that this magnitude-based approach maintains the attack's efficacy, as it accurately identifies critical weights, enabling a highly impactful attack. 
We observe that, without gradient information, setting \( \alpha = 0 \) allows the model to be attacked with only \(\mathbf{ 4.129 \times 10^{-9}} \%\) bit-flips, reducing model accuracy to zero. This degradation matches the attack's effectiveness at \( \alpha = 0.5 \), as depicted in Figure 6. 
Subsequent analysis of the attack vector confirms that identical weights and bits are selected for attack across both settings, underscoring AttentionBreaker's versatility and adaptability. This robustness is particularly significant as many defenses against BFAs assume that gradient access is essential for an attacker, prompting defenses to obscure or protect gradients. \emph{However, AttentionBreaker’s gradient-free capability bypasses these gradient-based defenses entirely by relying on magnitude-based importance, showing its potential to circumvent gradient-based restrictions. This adaptability highlights AttentionBreaker’s broad applicability in gradient-restricted environments, establishing it as a resilient approach for targeted model attacks.}

\subsubsection{AttentionBreaker Task Transferability}
\begin{figure}[t!]
  \centering
  \hfill
  \begin{subfigure}{0.48\linewidth}
    \centering
    \includegraphics[width=1\linewidth]{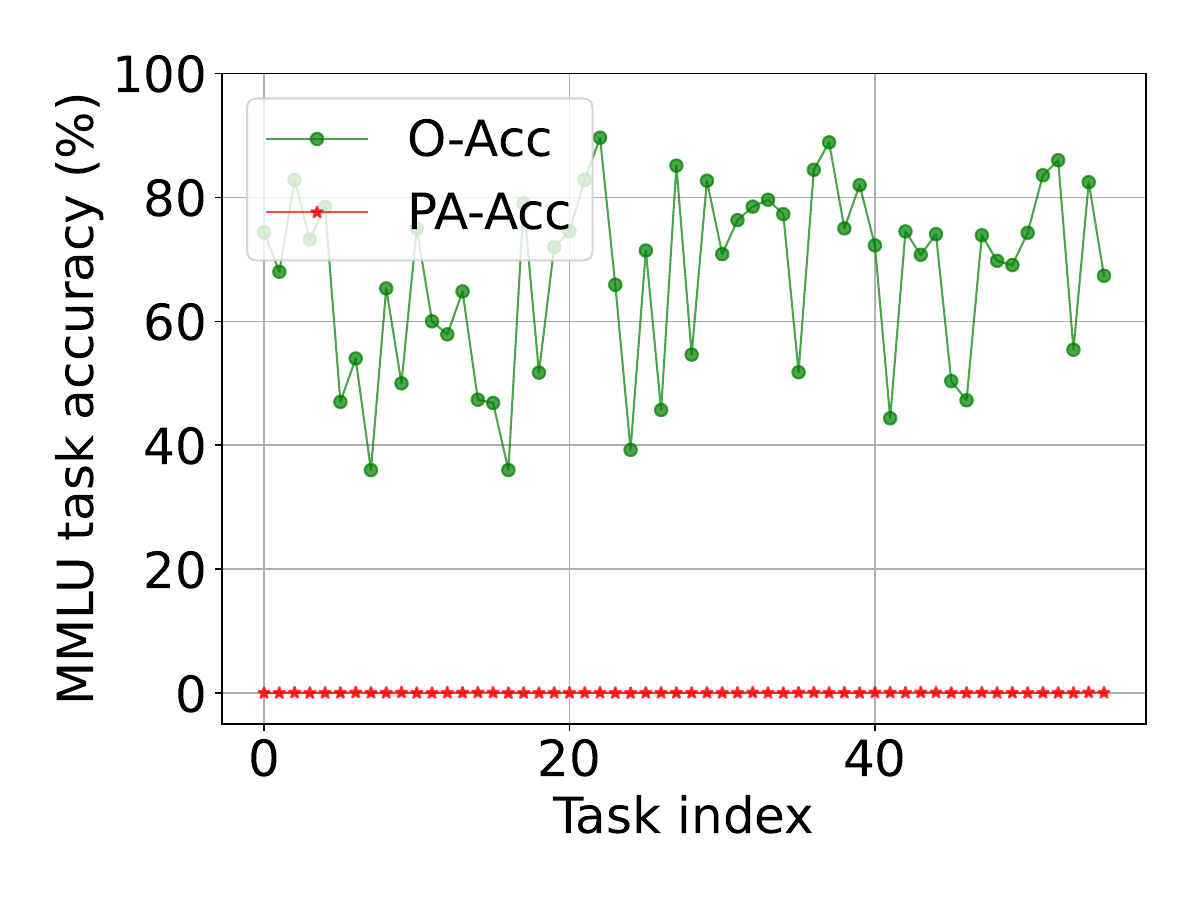}
  \caption{MMLU.}
  \label{fig:task_transferability}
  \end{subfigure}
  \begin{subfigure}{0.48\linewidth}
    \includegraphics[width=1\linewidth]{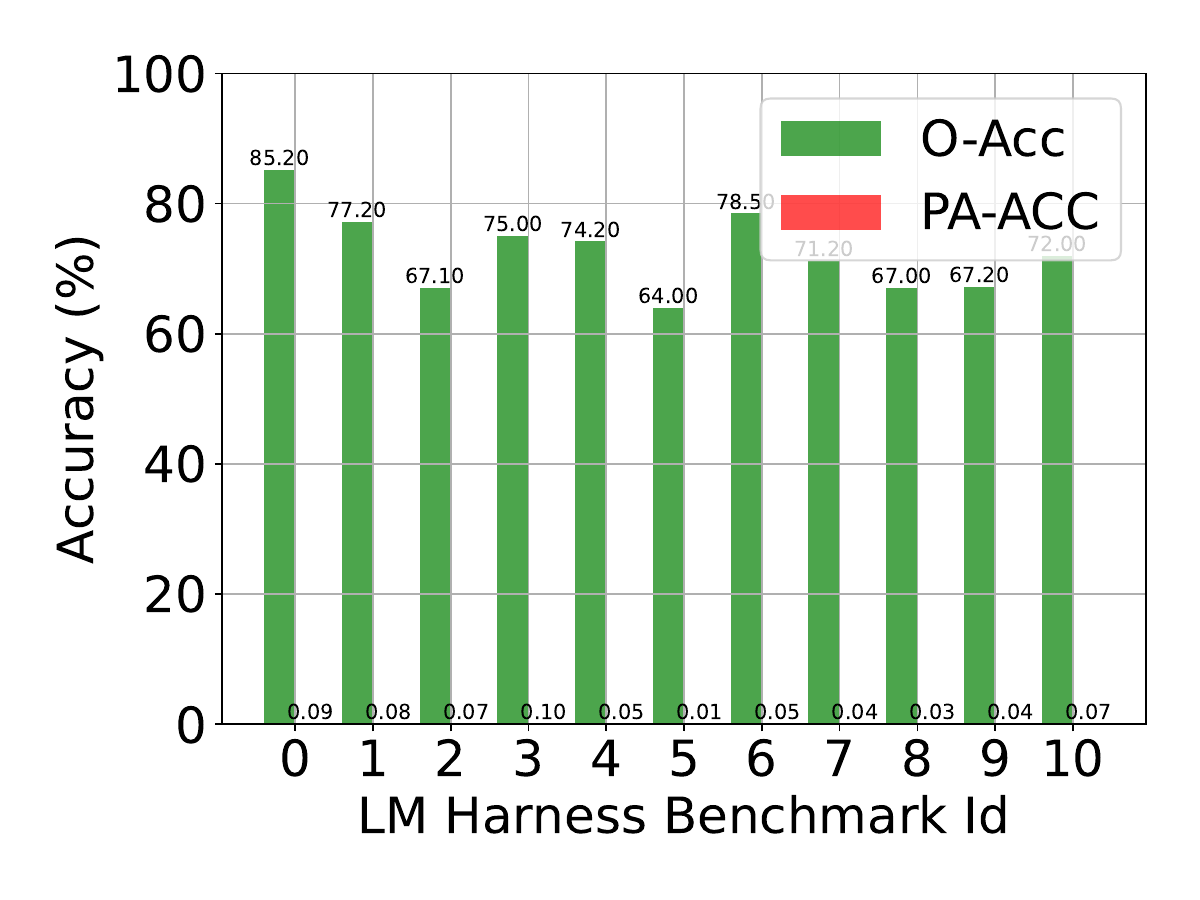}
  \caption{LM Harness.}
  \label{fig:task_transferability_common}
  \end{subfigure}
  \caption{LLaMA3-8B-Instruct W8 attack transferability analysis across various tasks.}
  \label{fig:transferability}
\end{figure}

This experiment examines the transferability of attack effects across distinct tasks. Specifically, it investigates whether an attack executed on Task A leads to observable adverse effects on Task B. If the attack effects indeed transfer between tasks, it would indicate a fundamental vulnerability in the model architecture, independent of the task at hand. Such a finding would underscore the severity of the model’s susceptibility to the attack. For this investigation, the LLaMA3-8B W8 model was subjected to a targeted attack on the MMLU task ``Astronomy''. Due to the attack, the accuracy of the model on ``Astronomy'' drops from the original accuracy (O-Acc) of 72.1\% to a post-attack accuracy (PA-Acc) of 0\%. The resulting impact was then systematically evaluated across additional MMLU tasks and other language benchmarks, including the LM harness benchmark. The results reveal a high degree of transferability, whereby the attack yields comparable degradation in performance across the evaluated tasks. This finding underscores the broad vulnerability of the model to such attacks. Figure \ref{fig:task_transferability} illustrates that an attack initially targeted at the ``Astronomy'' task leads to substantial accuracy reductions in other MMLU tasks. Similarly, an attack initially targeted at the ``AddSub'' task from the LM harness benchmark leads to severe model degradation on other LM harness tasks as well as captured in Figure \ref{fig:task_transferability_common}. \emph{This illustrates the pronounced transferability of the attack across varied tasks and domains.}


\subsubsection{AttentionBreaker Transferability to Fine-tuned models}
\begin{figure}[t!]
  \centering
  \begin{subfigure}{0.49\linewidth}
    \centering
    \includegraphics[width=1.\linewidth]{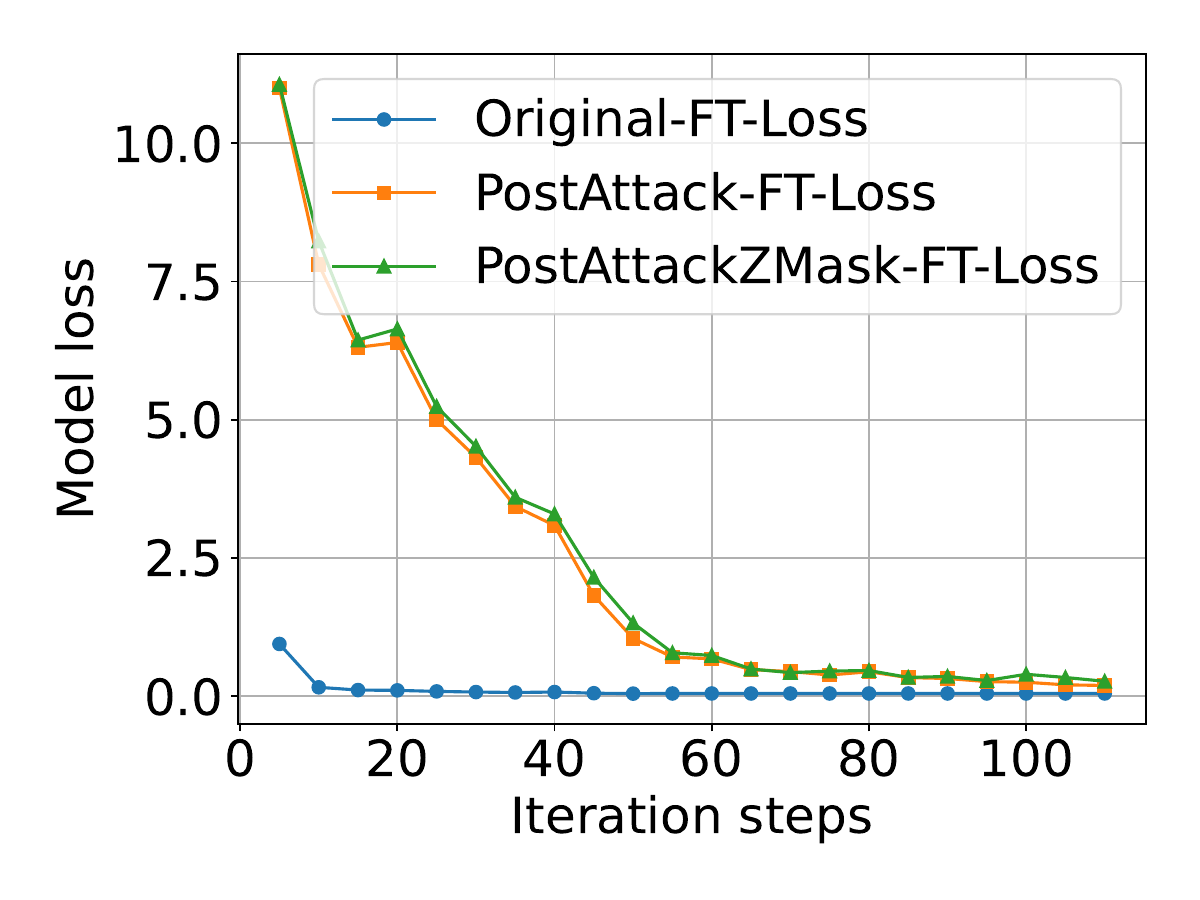}
  \caption{Fine-tuning model loss with iterations.}
  \label{fig:finetuning_loss}
  \end{subfigure}
  \begin{subfigure}{0.49\linewidth}
    \includegraphics[width=1.\linewidth]
{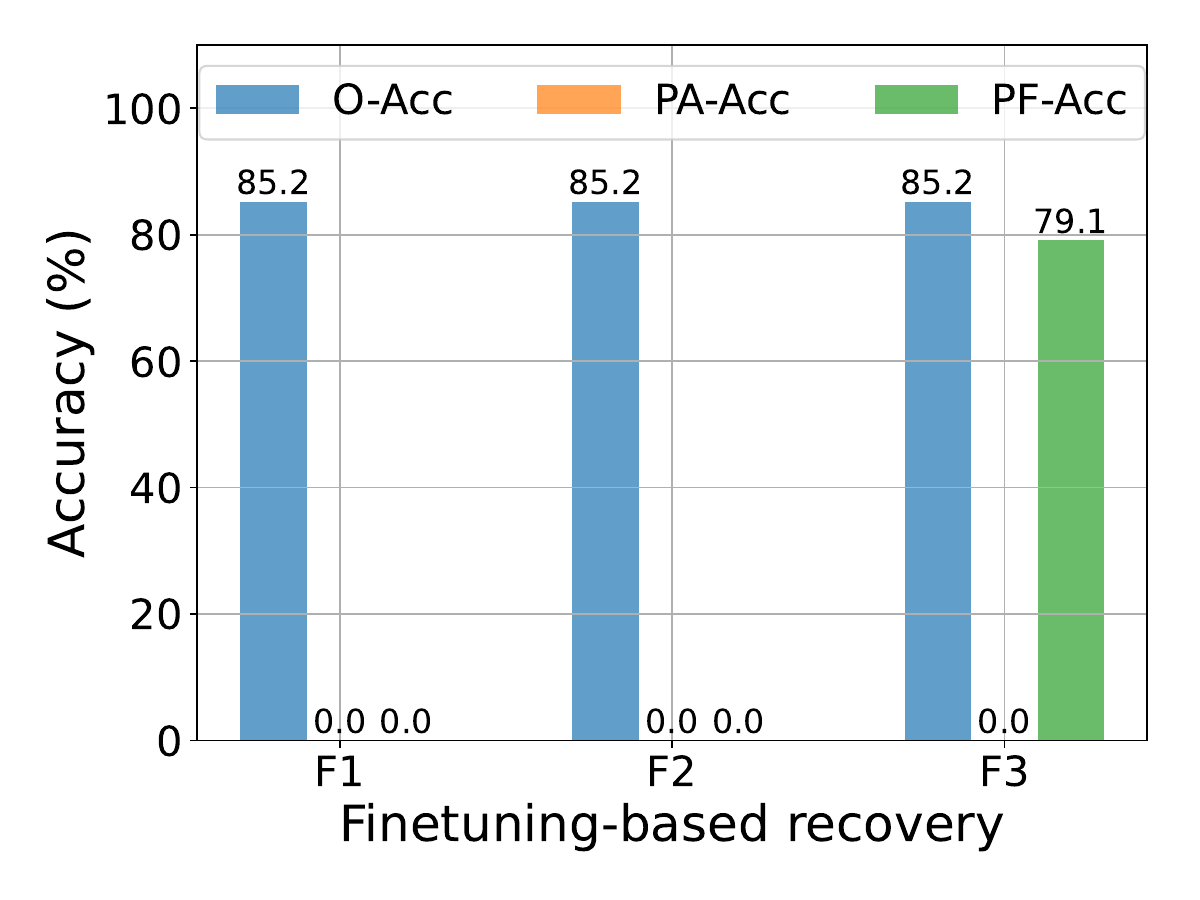}
    \caption{Fine-tuning-based recovery strategies.}
    \label{fig:finetuning_acc}
  \end{subfigure}
  \caption{Fine-tuning-based recovery after the attack: (a) validation loss with iteration steps and (b) model performance for different fine-tuning strategies. Metrics are calculated on the `AddSub' task of the LM Harness benchmark.}
  \label{fig:fine}
\end{figure}

This experiment evaluates the persistence of adversarial effects from an attack applied to a base model when, after an attack, the model is loaded with previously fine-tuned adapters for a specific task. Specifically, we fine-tune the LLaMA3-8B-Instruct W8 model on the ``AddSub'' task from the LM harness benchmark and save the resulting fine-tuned adapter. The fine-tuning progress is shown in Figure \ref{fig:finetuning_loss}, denoted as original fine-tuning or Original-FT-Loss. Next, we reload the original base model and perform the AttentionBreaker attack. As anticipated, the model exhibits a substantial loss increase and severe accuracy degradation, with accuracy plummeting from \(85.2\%\) to \(0\%\), as illustrated in Figure \ref{fig:finetuning_acc} (F1). Upon loading the fine-tuned adapter, we observe that the model accuracy remains at \(0\%\), indicating that \emph { the fine-tuned adapter alone is insufficient to restore model accuracy following the attack.}

\subsubsection{Fine-tuning-based Recovery}
This experiment assesses the persistence of adversarial effects introduced through the AttentionBreaker on a base model, specifically evaluating whether subsequent fine-tuning on a task can mitigate these impacts. We begin by attacking the LLaMA3-8B-Instruct W8 model, followed by evaluating the model’s accuracy on the ``AddSub'' task from the LM harness benchmark. As a result, the model’s accuracy drops drastically from an initial $85.2$\% to $0$\%, as depicted in Figure \ref{fig:finetuning_acc} (F2), indicating that the model output now consists largely of arbitrary strings devoid of task relevance.

Subsequently, we fine-tune the attacked model on the same task, conducting five epochs with $21$ iteration steps per epoch, as shown in Figure \ref{fig:finetuning_loss}. Throughout fine-tuning, the validation loss (or the post-attack fine-tuning loss PostAttack-FT-Loss) declines substantially from an initial $9.12$ to $0.27$, signaling effective task adaptation. Despite this apparent success in convergence, post-fine-tuning testing reveals that model accuracy remains at $0$\%, as shown in Figure \ref{fig:finetuning_acc}. This result suggests that while the model now outputs seemingly structured words rather than random strings, it fails to exhibit logical coherence or task-related reasoning, thus remaining unsuccessful in executing the logical deduction necessary for the ``AddSub'' task. \emph{The findings imply that fine-tuning does not suffice to restore task-specific functionality after the AttentionBreaker attack}.

\subsubsection{Zero-masked fine-tuning}
In this study, we investigate the persistence of adversarial effects introduced via the AttentionBreaker method on a base model and evaluate whether task-specific fine-tuning—applied after resetting the perturbed weights to zero—can effectively mitigate these impacts.
In particular, we aim to determine whether nullifying the impact of the bit-flipped weights (\textit{i.e.}, setting the perturbed weights to zero, \( w_i = 0 \)) permits effective recovery of model performance through fine-tuning on the same task. We begin attacking the LLaMA3-8B-Instruct W8 model using our AttentionBreaker framework, subsequently evaluating its accuracy on the ``AddSub'' task from the LM harness benchmark. Immediately following the attack, the model’s accuracy plunges from an initial $85.2\%$ to $0\%$, as illustrated in Figure \ref{fig:finetuning_acc}. 
Next, we \textcolor{black}{mask the attacked weights by resetting them to zero (\( w_i = 0 \))} and fine-tune the model on the same task, performing five epochs with 21 iterations per epoch. During fine-tuning, the validation loss (or PostAttackZMask-FT-Los) decreases significantly from an initial value of $9.03$ to $0.25$, as depicted in Figure \ref{fig:finetuning_loss}, indicating that the model is adapting effectively to the task requirements.

Upon fine-tuning, we test the model performance on the same task and observe that the model's accuracy is substantially recovered from $0\%$ to $79.1\%$, as shown in Figure \ref{fig:finetuning_acc} (F3). \emph{These findings suggest that fine-tuning, preceded by zeroing out perturbed weights, is sufficient to restore task-specific functionality, thereby mitigating the adversarial effects induced by the attack.}



\section{Ablation Study}\label{sec:ablation}

This section presents an ablation study to systematically evaluate and decide on specific components in our proposed method, such as $\alpha$, top-$n$, or which genetic algorithm to use for optimization.
\begin{figure}[t!]
  \centering
  \begin{subfigure}{0.46\linewidth}
    \includegraphics[width=1.1\linewidth]
{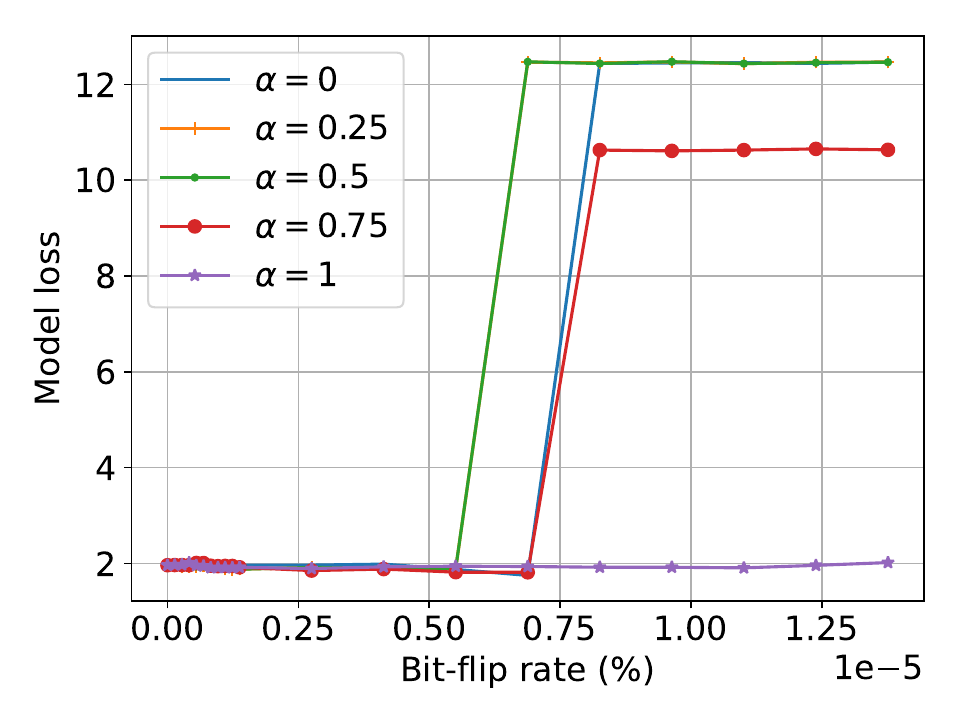}
    \caption{Sensitivity ratio ($\alpha$).}
    \label{fig:alphasweep}
  \end{subfigure}
  \hfill
  \begin{subfigure}{0.46\linewidth}
    \includegraphics[width=1.1\linewidth]
{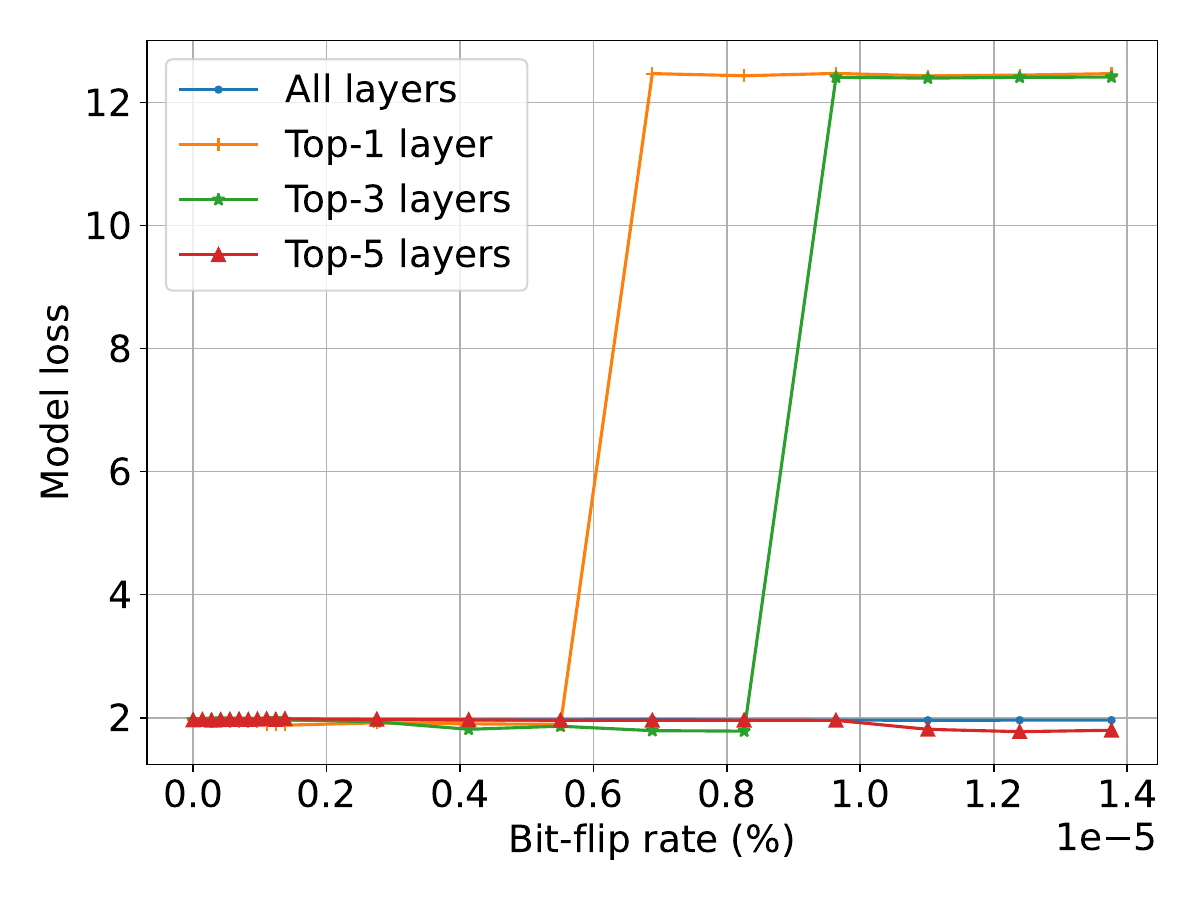}
    \caption{Top-ranked layer count.}
    \label{fig:2bitsweep}
  \end{subfigure}
  \caption{LLaMA3-8B-Instruct W8 ablation study on selecting weight subset. Analysis was performed using the `Astronomy' task of the MMLU benchmark.}
  \label{fig:llama38bit1}
\end{figure}

\begin{figure}[t!]
  \centering
  \begin{subfigure}{0.46\linewidth}
    \includegraphics[width=1.1\linewidth]
{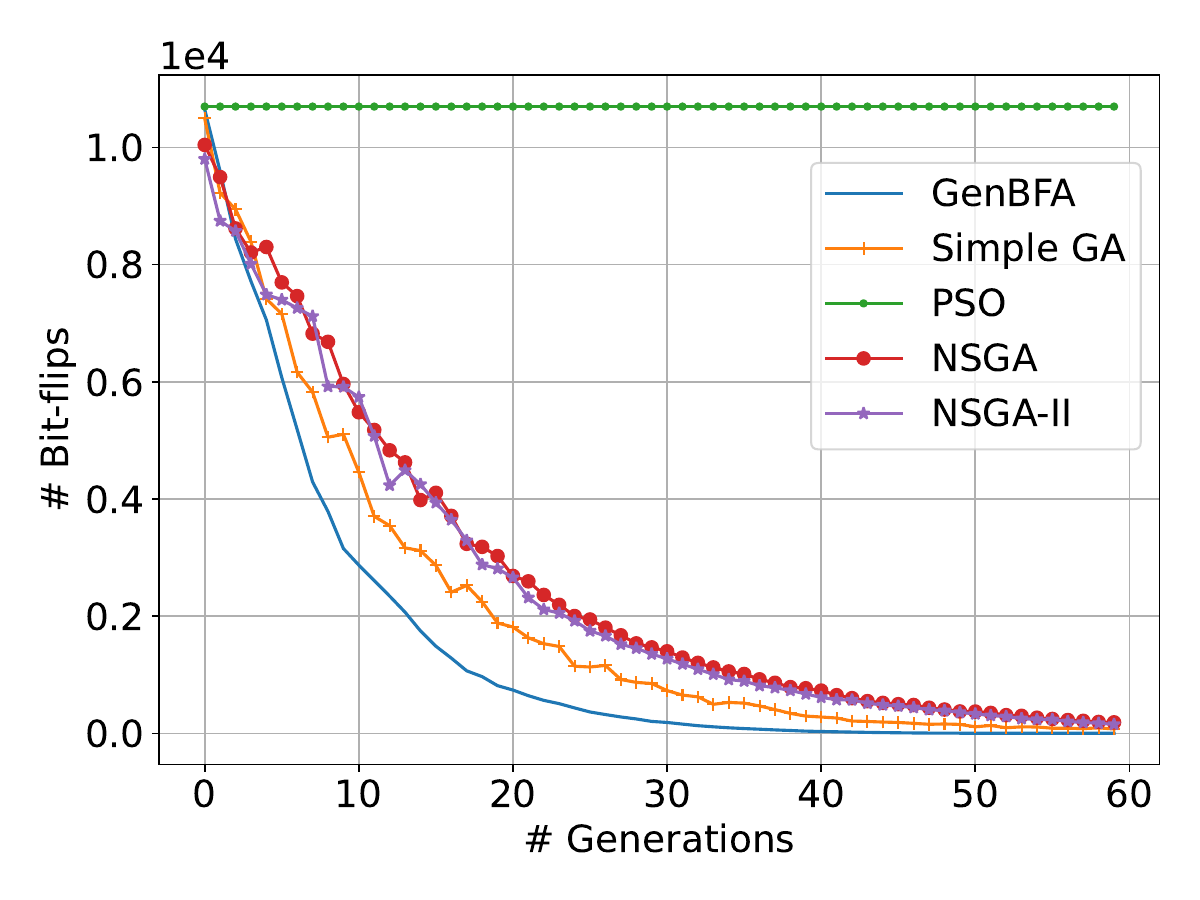}
    \caption{comparison between various genetic algorithms.}
    \label{fig:genetic_sweep}
  \end{subfigure}
  \hfill
  \begin{subfigure}{0.46\linewidth}
    \includegraphics[width=1.1\linewidth]
{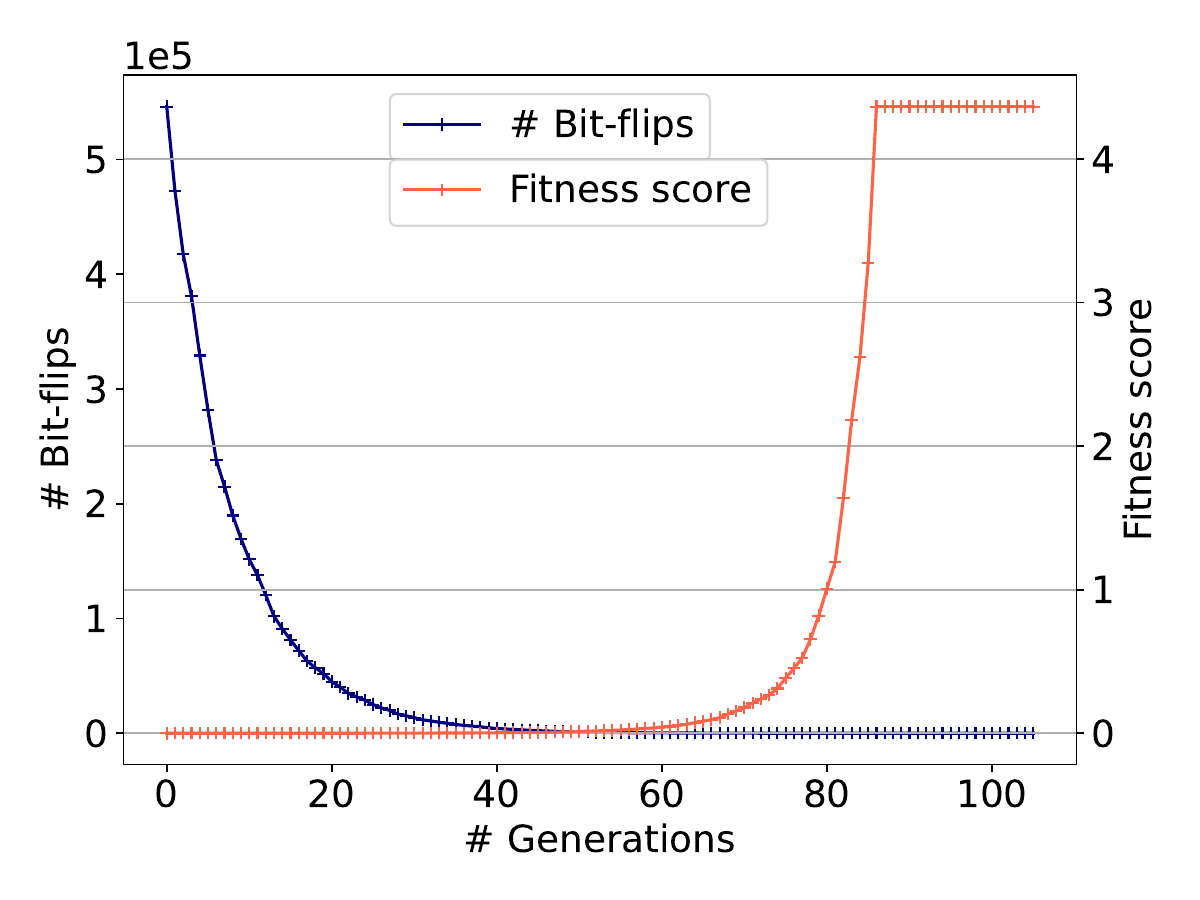}
    \caption{number of bit-flips and fitness score optimization.}
    \label{fig:genetic_sweep_gen}
  \end{subfigure}
  \hfill
  \caption{LLaMA3-8B-Instruct W8 ablation study on genetic optimization to select critical weights. Fitness scores for the genetic algorithms are calculated on ``Astronomy'' task of MMLU benchmark.}
  \label{fig:llama38bit2}
\end{figure}

\subsection{Sensitivity Ratio Variation}
We vary $\alpha$ between $|\nabla\mathbf{W}|$ and $|\mathbf{W}|$ in the min-max normalized importance score calculation as described in Section \ref{sec:layer_sensitivity}. \textcolor{black}{Under different values of $\alpha$, the model loss at different bit-flip rate is depicted in Figure \ref{fig:alphasweep}. As observed, an optimal value of \(\alpha = 0.5\) achieves the best results, closely followed by \(\alpha = 0\) and \(\alpha = 0.75\). \emph{Consequently, \(\alpha = 0.5\) is selected for subsequent experiments.}}

\subsection{All vs top-$n$ Layer Sensitivity Analysis}
In this experiment, we investigate the impact of varying the selection of top-\( n \) layers on model performance by comparing the effects of sub-sampling from the top-\( n \) layers versus sub-sampling from all layers. With a fixed sensitivity parameter of \( \alpha = 0.5 \), we increment \( n \) from 1 to 5. The results demonstrate that sub-sampling from the top-\( n \) layers leads to a higher loss than sub-sampling from all layers, as illustrated in Figure \ref{fig:2bitsweep}. Among the top-\( n \) layers, selecting the top-1 layers yields the best performance, followed by top-3 and top-5. The weight subset selection and optimization based on top-\( n \) layers provide superior results, as the constrained weight subspace facilitates faster convergence. \emph{Consequently, we fix \( n = 1 \) for all subsequent experiments.} 


\subsection{Genetic Algorithms Variation}

We extensively evaluate the performance of our proposed evolutionary optimization technique against established algorithms, including the Simple Genetic Algorithm (SGA) \cite{vose1999simple}, Particle Swarm Optimization (PSO) \cite{eberhart1995new}, Non-Dominated Sorting Genetic Algorithm (NSGA) \cite{6791727}, and NSGA-II~\cite{deb2000fast, lambora2019genetic}. To ensure comparability, we standardize parameters across all algorithms, setting the population size to 100, mutation rate to 0.1, and crossover rate to 0.9. Additionally, we apply modifications in population initialization, crossover, and mutation strategies as in our custom implementation. Results in Figure \ref{fig:genetic_sweep} demonstrate that our modified genetic algorithm achieves superior convergence speed, owing to optimizations tailored to problem-specific requirements. \emph{Consequently, this optimized genetic algorithm is employed in all subsequent experiments.}
Figure \ref{fig:genetic_sweep_gen} illustrates the optimization of the weight set within AttentionBreaker’s genetic implementation. Notably, the required weight count decreases exponentially as the fitness score rises, reflecting rapid improvement. After approximately 80–90 iterations, both metrics plateau, indicating convergence; at this stage, the optimal solution is returned as the output. 


\section{Discussion}\label{sec:defence}
As LLMs become increasingly pervasive in various aspects of daily life, their integration into numerous applications often occurs without users' explicit awareness. Consequently, any adversarial attack that undermines the integrity of these models poses significant risks in mission-critical scenarios. Thus, it is imperative to extensively study and explore LLM vulnerabilities to enhance the security and reliability of this technology. To counter such attacks exemplified by the proposed AttentionBreaker, we suggest an adaptive defense mechanism to disrupt the attack's underlying strategy. This defense, inspired by XOR-cipher logic locking~\cite{kamali2022advances}, preemptively identifies a critical set of vulnerable model parameters. A subset of these parameters is then encrypted using an XOR-cipher. When provided with the correct key set, the model maintains its original accuracy during inference; conversely, using an incorrect key leads to significant degradation in accuracy. Therefore, an attacker lacking knowledge of the key will encounter a network with substantially reduced baseline accuracy, rendering further attacks impractical from an adversarial standpoint.
\section{Conclusion}\label{sec:conclusion}
In this work, we introduced a novel adversarial BFA framework, AttentionBreaker, for transformer-based models such as LLMs. For the first
time, AttentionBreaker applies a novel sensitivity analysis along with evolutionary optimization to guide adversarial attacks on LLMs. The AttentionBreaker reveals the extreme vulnerability of
LLMs to adversarial BFAs. Specifically, perturbing merely $\mathbf{3}$ bit-flips $\mathbf{4.129 \times 10^{-9}}\%$ of the parameters) in LLaMA3 results in catastrophic performance degradation, with MMLU scores dropping from $67.3$\% to $0$\%
and perplexity increased sharply from 12.6 to \(4.72\times10^5\). These findings underscore the
effectiveness of our method in exposing the vulnerability of LLMs to such adversarial interventions,
highlighting the need for robust defense mechanisms in future model designs.

\balance

\bibliographystyle{IEEEtran}
\bibliography{ref}
\end{document}